\documentclass[useAMS,usenatbib]{mn2e}
\usepackage{amsmath}
\usepackage{amssymb}
\usepackage{graphicx}
\usepackage{epsfig}
\makeatletter{}\def\LCDM{\mbox{$\Lambda$CDM }}
\def\Mpch{\mbox{$h^{-1}$Mpc}}
\def\Gpch{\mbox{$h^{-1}$Gpc}}

\def\M200{\mbox{$M_{\rm 200 }$}}

\def\R200{\mbox{$R_{\rm 200 }$}}

\def\V200{\mbox{$V_{\rm 200 }$}}
\def\kms{\mbox{$\rm {km\,sec}^{-1}$ }}
\def\Nbody{$N$-body}

\newcommand{\vc}[1]{\ensuremath{\bf #1}}
\renewcommand{\vec}[1]{ {\bmath #1} }
\newcommand{\Ng}{\mbox{$N_{\rm g}$}}
\newcommand{\Np}{\mbox{$N_{\rm p}$}}
\newcommand{\Ns}{\mbox{$N_{\rm s}$}}
\newcommand{\lsim}{\mbox{${\,\hbox{\hbox{$ < $}\kern -0.8em \lower 1.0ex\hbox{$\sim$}}\,}$}}
\newcommand{\gsim}{\mbox{${\,\hbox{\hbox{$ > $}\kern -0.8em \lower 1.0ex\hbox{$\sim$}}\,}$}}

\def\beqn{\vspace{2mm}
\begin{eqnarray}} 
\def\eeqn{\vspaceg{2mm} 
\end{eqnarray}}

\newcommand{\be}{\begin{equation}}
\newcommand{\ee}{\end{equation}}
\newcommand{\ba}{\begin{eqnarray}}
\newcommand{\ea}{\end{eqnarray}}
\newcommand{\brr}{\begin{array}}
 
\newcommand{\err}{\end{array}}
\newcommand{\bc}{\begin{center}}
\newcommand{\ec}{\end{center}}

\title[Dark matter statistics for large galaxy catalogs]{Dark matter statistics for large galaxy
catalogs: power spectra and  covariance matrices}

\makeatletter{}\author[A.~Klypin \& F.~Prada]
  {Anatoly~Klypin$^{1}$ and Francisco~Prada
   \vspace{0.2cm}\\ 
  $^1$Astronomy Department, New Mexico State University, Las Cruces, NM, USA\\
}

\author[Klypin \& Prada]{Anatoly~Klypin$^1$\thanks{E-mail: aklypin@nmsu.edu} and Francisco~Prada$^{2}$ \\
 \vspace{-0.2cm}\\
$^{1}$ Astronomy Department, New Mexico State University, Las Cruces, NM, USA\\
$^2$ Instituto de Astrof\'{\i}sica de Andaluc\'{\i}a (CSIC), Glorieta de 
     la Astronom\'{\i}a, E-18080 Granada, Spain \\
\\
}

\begin{document}
\maketitle
\label{firstpage}
\begin{abstract}
  Upcoming and existing large-scale surveys of galaxies require
  accurate theoretical predictions of the dark matter clustering statistics for
  thousands of mock galaxy catalogs. We demonstrate that this goal can be
  achieve with our new Parallel Particle-Mesh (PM) \Nbody\ code (PPM-GLAM) at a very low
  computational cost. We run about 15,000 simulations
  with $\sim 2$\,billion particles that provide $\sim 1$\%
  accuracy of the dark matter power spectra $P(k)$ for wave-numbers up
  to $k\sim 1h{\rm Mpc}^{-1}$.  Using this large data-set we study the
  power spectrum covariance matrix, the stepping stone for producing
  mock catalogs. In contrast to many previous analytical and numerical
  results, we find that the covariance matrix normalised to the power
  spectrum $C(k,k^\prime)/P(k)P(k^\prime)$ has a complex structure
  of non-diagonal components. It has an upturn at small $k$, followed by a minimum at
  $k\approx 0.1-0.2\,h{\rm Mpc}^{-1}$. It also has a maximum at
  $k\approx 0.5-0.6\,h{\rm Mpc}^{-1}$. The normalised covariance
  matrix strongly evolves with redshift:
  $C(k,k^\prime)\propto \delta^\alpha(t)P(k)P(k^\prime)$, where
  $\delta$ is the linear growth factor and $\alpha \approx 1-1.25$, which
  indicates that the covariance matrix depends on cosmological
  parameters. We also show that waves longer than $1\Gpch$ have very
  little impact on the power spectrum and covariance matrix. This
  significantly reduces the computational costs and complexity of
  theoretical predictions: relatively small volume $\sim (1\Gpch)^3$ simulations
  capture the necessary properties of dark matter clustering statistics. All the power spectra 
  obtained from many thousands of our simulations are  publicly available.

\end{abstract}


\makeatletter{}\section{Introduction}

Accurate theoretical predictions for the clustering properties of different
galaxy populations are crucial for the success of massive
observational surveys such as SDSS-III/BOSS \citep[e.g.,][]{BOSS2016}, SDSS-IV/eBOSS
\citep{eBOSS}, DESI \citep{DESI2016} and Euclid \citep{Euclid} that
have or will be able to measure the positions of millions of
galaxies. The predictions of the statistical properties of the distribution of
galaxies expected in modern cosmological models involve a number of
steps: it starts with the generation of the dark matter clustering and proceeds with
placing ``galaxies'' according to some bias prescriptions
\citep[see][for a review]{Chuang2015,Monaco2016}.

While a substantial effort has been made to estimate the non-linear
matter power spectrum $P(k)$ \citep[e.g.,][]{Smith2003,Heitmann2009,Schneider2016}, the
covariance matrix is a much less studied quantity that is required for
the analysis of the observational data.  Most of the time the main
attention shifts to the last step, i.e., the placement of ``galaxies'' in the density field
with all the defects and uncertainties of estimates of the dark matter
distribution and velocities being often incorporated into the biasing
scheme \citep[e.g.,][]{Chuang2015}.  However, one needs to accurately
reproduce and to demonstrate that whatever algorithm or prescription
is adopted to mock galaxies, also is able to match the clustering and
covariance matrix of the dark matter field.

The covariance matrix $C(k,k^\prime)$ is the second-order statistics
of the power spectrum:
$C(k,k^\prime) =\langle P(k)P(k^\prime)\rangle - \langle P(k)\rangle
\langle P(k^\prime)\rangle$,
where $\langle \dots \rangle$ implies averaging over an ensamble of
realizations. 
The power spectrum covariance and its cousin the covariance of the
correlation function play an important role in estimates of the
accuracy of measured power spectrum, and the inverse covariance
matrices are used in estimates of cosmological parameters deduced from
these measurements
\citep[e.g.,][]{Anderson2012,Sanchez2012,Dodelson2013,Percival2014}. The power
spectrum covariance matrix measures the degree of non-linearity and
mode coupling of waves with different wave-numbers. As such it is
an interesting entity on its own.

Because it is expensive to produce thousands of simulations
\citep[e.g.,][]{Taylor2013,Percival2014} using standard
high-resolution $N$-body codes, there were relatively few publications
on the structure and evolution of the covariance matrix. Most of the
results are based on simulations with relatively small number of
particles (16~million as compared with our $\sim 2$~billion) and small
computational volumes of $500-600\,\Mpch$
\citep{Takahashi2009,Li2014,Blot2015}. There was no systematic
analysis of the effects of mass, force resolution and volume on the
covariance matrix.

So far, there are some uncertainties and disagreements even on the
shape of the covariance matrix.
\citet{Neyrinck2011,Mohammed2014,Carron2015} argue that once the
Gaussian diagonal terms are removed, the covariance function should be
a constant. This disagrees with numerical \citep{Li2014,Blot2015} and
analytical \citep{Bertolini2016,Mohammed2016} results that indicate
that the non-diagonal components of the covariance matrix have quite
complex structure.

There are also issues and questions related with numerical
simulations. How many realizations are needed for accurate
measurements of the covariance matrix \citep{Taylor2013,Percival2014}
and how can it be reduced \citep{Pope2008,Pearson2016,Joachimi2017}?
What resolution and how many time-steps are needed for accurate
estimates? How large the simulation volume should be to adequately
probe relevant scales and to avoid defects related with the box size
\citep{Gnedin2011,Li2014}?

The goal of our paper is to systematically study the structure of the
matter covariance matrix using a large set of \Nbody\ simulations.
The development of $N$-body cosmological algorithms has been an active
field of research for decades.  However, the requirements for the
generation of many thousands of high-quality simulations are
extreme. Existing codes such as {\sc GADGET}, {\sc RAMSES}, or {\sc
  ART} are powerful for large high-resolution simulations, but they
are not fast enough for medium quality large-number of realisations
required for analysis and interpretation of large galaxy surveys. New
types of codes \citep[e.g.,][]{QPM,COLA,Feng2016} are being developed
for this purpose. Here, we present the performance results of our new
\Nbody\ Parallel Particle-Mesh GLAM code (PPM-GLAM), which is the core
of the GLAM (GaLAxy Mocks) pipeline for the massive production of
large galaxy catalogs. PPM-GLAM generates the density field, including
peculiar velocities, for a particular cosmological model and initial
conditions.

The optimal box size of the simulations is of particular interest for producing
thousands of mock galaxy catalogs and for the studies of large-scale galaxy clustering.  With
the upcoming observational samples tending to cover larger observational
volumes ($\sim 50\,{\rm Gpc}^3$), one naively expects that the computational volumes should be
large enough to cover the entire observational sample. Indeed, this will be
the ideal case, if we were to measure some statistics that involve waves as
long as the whole sample. The problem with producing extra large
simulations is their computational cost. For accurate predictions one needs to
maintain the resolution on small scales. With a fixed resolution the
cost of computations scales with volume. So, it increases very
quickly and becomes prohibitively expensive.

However, most of the observable clustering statistics rely on relatively small
scales, and thus would not require extra large simulation boxes. For
example, the Baryonic Acoustic Oscillations (BAO) are on scales of
$\sim 100\,\Mpch$ and  the power spectrum of galaxy clustering is
typically measured on wave-numbers $k\gsim 0.05\,h{\rm Mpc}^{-1}$. In
order to make accurate theoretical predictions for these scales, we
need relatively small $\sim 1\,\Gpch$ simulation volumes. 

What will be the consequences of using small computational volumes?
One may think about few. For example, long-waves missed in small
simulation boxes may couple with small-scale waves and produce larger
power spectrum (and potentially covariance matrix). This is the so
called Super Sample Covariance
\citep[SSC,][]{Gnedin2011,Li2014,Wagner2015,Baldauf2016}. Another
effect is the statistics of waves. By replicating and stacking small
boxes (to cover large observational volumes) we do not add the
statistics of small waves, which we are interested in. For example,
the number of pairs with separation of say $100\,\Mpch$ will be
defined only by how many independent pairs are in a small box, and not
by the much larger number of pairs found in large observational
samples. These concerns are valid, but can be resolved in a number of
ways. SSC effects depend on the volume and become very small as the
computational volume increases. Statistics of waves can be re-scaled
proportionally to the volume
\citep[e.g.,][]{Mohammed2014,Bertolini2016}. We investigate these
issues in detail in our paper, together with the performance of the
power spectrum depending on the numerical parameters of our PPM-GLAM
simulations. All the power spectra obtained from many thousands of our
simulations are made publicly available.

In Section~\ref{sec:ppmglam} we discuss the main features of our PPM-GLAM
simulation code. More detailed description and tests are presented in
the Appendixes. The  suite of simulations used in this paper is presented in
Section~3.  Convergence and accuracy of the power spectrum are
discussed in Section~4. The results on the covariance matrix of the
power spectrum are given in Section~5.  We campare our results with
other works in Section~6. Summary and discussion of our results are
presented in Section~7.

\makeatletter{}\begin{table*}
 \begin{minipage}{16.cm}
\caption{Numerical and cosmological parameters of different simulations.
  The columns give the simulation identifier, 
  the size of the simulated box in $h^{-1}\,{\rm Mpc}$,
  the number of particles, 
  the mass per simulation particle $m_p$ in units $h^{-1}\,M_\odot$, the mesh size $\Ng^3$,
  the  gravitational softening length $\epsilon$ in units of $h^{-1}\,{\rm Mpc}$, the number of time-steps $N_s$, the amplitude of perturbations $\sigma_8$, the matter density $\Omega_m$,
  the number of realisations $N_r$ and the total volume in $[h^{-1}\,{\rm Gpc}]^3$}
\begin{tabular}{ l | c | c | c |  c|  c | c | c | c | r |r }
\hline  
Simulation & Box &  particles  & $m_p$                    & $\Ng^3$  & $\epsilon$ & $N_{\rm s}$ & $\sigma_8$ & $\Omega_m$ & $N_r$ & Total Volume
\tabularnewline
  \hline 
PPM-GLAM A0.5         & 500$^3$    & 1200$^3$ & $6.16\times 10^9$   & 2400$^3$ & 0.208 & 181 & 0.822 & 0.307 & 680 & 85
\tabularnewline
PPM-GLAM A0.9         & 960$^3$    & 1200$^3$ & $4.46\times 10^{10}$ & 2400$^3$ & 0.400 & 136 & 0.822 & 0.307  & 2532 & 2240
\tabularnewline
PPM-GLAM A1.5         & 1500$^3$   & 1200$^3$ & $1.66\times 10^{11}$ & 2400$^3$ & 0.625 & 136 & 0.822 & 0.307  & 4513 & 15230
\tabularnewline
PPM-GLAM A2.5         & 2500$^3$   & 1000$^3$ & $1.33\times 10^{12}$ & 2000$^3$ & 1.250 & 136 & 0.822 & 0.307  & 1960 & 30620
\tabularnewline
PPM-GLAM A4.0         & 4000$^3$   & 1000$^3$ & $5.45\times 10^{12}$ & 2000$^3$ & 1.250 & 136 & 0.822 & 0.307  & 4575 & 292800
\tabularnewline
PPM-GLAM B1.0a        & 1000$^3$   & 1600$^3$ & $2.08\times 10^{10}$  & 3200$^3$ & 0.312 & 147 & 0.828 & 0.307  & 10 & 10
\tabularnewline
PPM-GLAM B1.0b        & 1000$^3$   & 1300$^3$ & $1.78\times 10^{10}$ & 2600$^3$ & 0.385 & 131 & 0.828 & 0.307  & 10 & 10
\tabularnewline
PPM-GLAM B1.5         & 1500$^3$   & 1300$^3$ & $3.88\times 10^{10}$ & 2600$^3$ & 0.577 & 131 & 0.828 & 0.307  & 10 & 33
\tabularnewline
PPM-GLAM C1a         & 1000$^3$   & 1000$^3$ & $8.71\times 10^{10}$ & 3000$^3$ & 0.333 & 302 & 0.828 & 0.307  & 1 & 1
\tabularnewline
PPM-GLAM C1b         & 1000$^3$   & 1000$^3$ & $8.71\times 10^{10}$ & 4000$^3$ & 0.250 & 136 & 0.828 & 0.307  & 1 & 1
\tabularnewline
\hline
BigMDPL$^1$    & 2500$^3$ & 3840$^3$ & $2.4 \times 10^{10}$  & --      & 0.010 & --  & 0.828 & 0.307   & 1 & 15.6 
\tabularnewline
HMDPL$^1$        & 4000$^3$ & 4096$^3$ & $7.9 \times 10^{10}$  & --      & 0.025 &  -- & 0.828 & 0.307   & 1 & 64
\tabularnewline
\hline
Takahashi et al.$^2$ & 1000$^3$ & 256$^3$ & $1.8 \times 10^{12}$  & --      & -- &  -- & 0.760 & 0.238   & 5000 & 5000
\tabularnewline
Li et al.$^3$ & 500$^3$ & 256$^3$ & $2.7 \times 10^{11}$  & --      & -- &  -- & 0.820 & 0.286   & 3584 & 448
\tabularnewline
Blot et al.$^4$ & 656$^3$ & 256$^3$ & $1.2 \times 10^{12}$  & --      & -- &  -- & 0.801 & 0.257   & 12288 & 3469
\tabularnewline
BOSS QPM$^5$           & 2560$^3$ & 1280$^3$ & $3.0 \times 10^{11}$  & 1280$^3$  & 2.00 &  7 & 0.800 & 0.29   & 1000 & 16700
\tabularnewline
WiggleZ COLA$^6$   & 600$^3$ & 1296$^3$ & $7.5 \times 10^{9}$ & $3888^3$   & -- & 10 & 0.812 & 0.273   & 3600 & 778
\tabularnewline
\hline
\multicolumn{11}{l}{\quad {\it References:} $^1$\citet{Klypin2016}, $^2$\citet{Takahashi2009}, 
     $^3$\citet{Li2014}, $^4$\citet{Blot2015}, }
\tabularnewline
\multicolumn{11}{l}{\quad {\it \phantom{References:}}    $^5$\citet{QPM},  $^6$\citet{Koda2016}       }          
\tabularnewline

\end{tabular}
\label{table:simtable}
\vspace{-5mm}
\end{minipage}
\end{table*}

\makeatletter{}\section{Parallel Particle-Mesh GLAM code} 
\label{sec:ppmglam}

There are a number of advantages of cosmological Particle-Mesh (PM)
codes \citep{Klypin1983,HockneyEastwood,PM1997} that make them useful
on their own to generating a large number of galaxy mocks \citep[e.g.,
QPM, COLA, FastPM;][]{QPM,COLA,Feng2016}, or as a component of more
complex hybrid TREE-PM \citep[e.g., Gadget2,
HACC;][]{Gadget2,HACC} and Adaptive-Mesh-Refinement codes
\citep[e.g. ART, RAMSES, ENZO;][]{ART,RAMSES,ENZO2014}. Cosmological
PM codes are the fastest codes available and they are simple.

We have developed and thoroughly tested a new parallel version of the
Particle-Mesh cosmological code, that provides us with a tool to
quickly generate a large number of \Nbody\ cosmological simulations
with a reasonable speed and acceptable resolution. We call our code
Parallel Particle-Mesh GLAM (PPM-GLAM), which is the core of the GLAM
(GaLAxy Mocks) pipeline for massive production of galaxy
catalogs. PPM-GLAM generates the density field, including peculiar
velocities, for a particular cosmological model and initial
conditions. Appendix~\ref{sec:AppendixA} gives the details of the code
and provides tests for the effects of mass and force resolutions, and
the effects of time-stepping. Here, we discuss the main features of
the PPM-GLAM code and provide the motivation for the selection of
appropriate numerical parameters.

The code uses a regularly spaced three-dimensional mesh of size
$N_{\rm g}^3$ that covers the cubic domain $L^3$ of a simulation box.
The size of a cell $\Delta x =L/N_{\rm g}$ and the mass of each
particle $m_{\rm p}$ define the force and mass resolution
respectively:
\begin{eqnarray}
m_{\rm p} &=& \Omega_m \, \rho_{\rm cr,0}\left[\frac{L}{N_{\rm p}}\right]^3 = \\
         &=& 8.517\times 10^{10}\left[\frac{\Omega_m}{0.30}\right]
         \left[\frac{L/\Gpch}{N_{\rm p}/1000}\right]^3h^{-1}M_\odot,\\
         \Delta x &=& \left[\frac{L/\Gpch}{N_{\rm g}/1000}\right]\Mpch,
\end{eqnarray}
where $N_{\rm p}^3$ is the number of particles and $\rho_{\rm cr,0}$
is the critical density of the universe at present.

PPM-GLAM solves the Poisson equation for the gravitational potential
in a periodical cube using a Fast Fourier Transformation (FFT)
algorithm. The dark matter density field used in the Poisson equation
is obtained with the Cloud-In-Cell (CIC) scheme using the positions of
dark matter particles. Once the gravitational potential is obtained,
it is numerically differentiated and interpolated to the position of
each particle. Then, particle positions and velocities are advanced in
time using the second order leap-frog scheme. The time-step is
increased periodically as discussed in Appendix~A.  Thus, a
standard PM code has three steps that are repeated many times until
the system reached its final moment of evolution: (1) Obtain the
density field on a 3D-mesh that covers the computational volume, (2)
Solve the Poisson equation, and (3) Advance particles to the next
moment of time.

The computational cost of a single PPM-GLAM simulation depends on the number of time-steps
$N_s$, the size of the 3D-mesh $N_{\rm g}^3$, and the adopted number of particles
$N_{\rm p}^3$. The CPU required to solve the Poisson equation is mostly
determined by the cost of performing a single 1D-FFT.   We incorporate
all numerical factors  into one
coefficient and write the CPU for the Poisson solver as $AN_{\rm g}^3$. The
costs of density assignment and particle displacement (including
potential differentiation) scale proportionally to $N_{\rm p}^3$. In total,
the CPU time $T_{\rm tot}$ required for a single PPM-GLAM run is:
\begin{equation}
T_{\rm tot} = \Ns\left[A\Ng^3 + (B+C) \Np^3 \right],
\label{eq:cpu}
\end{equation}
where $B$ and $C$ are the coefficients for scaling the CPU estimate
for particle displacements and density assignment. These numerical
factors were estimated for different processors currently used for
$N$-body simulations and are given in Table~\ref{table:cpu}.
For a typical simulation analysed in this
paper ($\Ng=2400$, $\Np=\Ng/2$) the CPU per time-step is
$\sim 0.5$\,hours and wall-clock time per step $\sim 1-3$\,minutes.
The total cost of 1000 PPM-GLAM realizations with $\Ns=150$ is 75K CPU hours, which
is a modest allocation even for a small computational cluster or a
supercomputer center.

Memory is another critical factor that should be considered when selecting
the parameters of our simulations. PPM-GLAM uses only one 3D-mesh for storing
both density and gravitational potential, and only one set of particle coordinates
and velocities. Thus, for single precision variables the total required memory $M_{\rm tot}$ is:

\begin{eqnarray}
M_{\rm tot} &=& 4N_{\rm g}^3 + 24N_{\rm p}^3\,\, {\rm Bytes}, \\
  &=& 29.8\left(\frac{\Ng}{2000}\right)^3+
22.3\left(\frac{\Np}{1000}\right)^3{\rm GB}, \\
&=&52\left(\frac{\Np}{1000}\right)^3{\rm GB}, \,\, {\rm for}\quad\Ng=2\Np.
\label{eq:mem}
\end{eqnarray}

\makeatletter{}\begin{figure*}
\centering
\includegraphics[width=0.495\textwidth]
{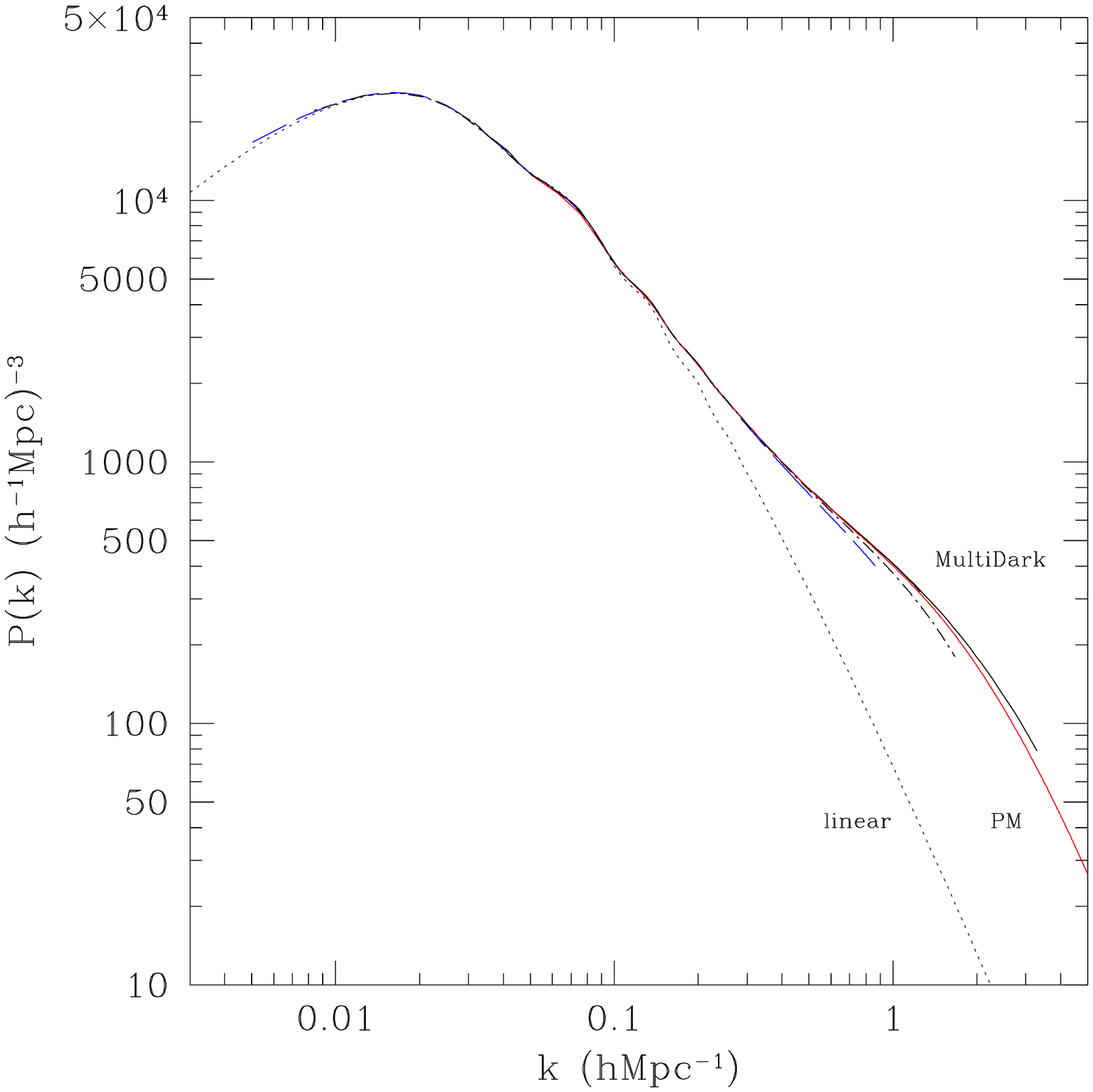}
\includegraphics[width=0.495\textwidth]
{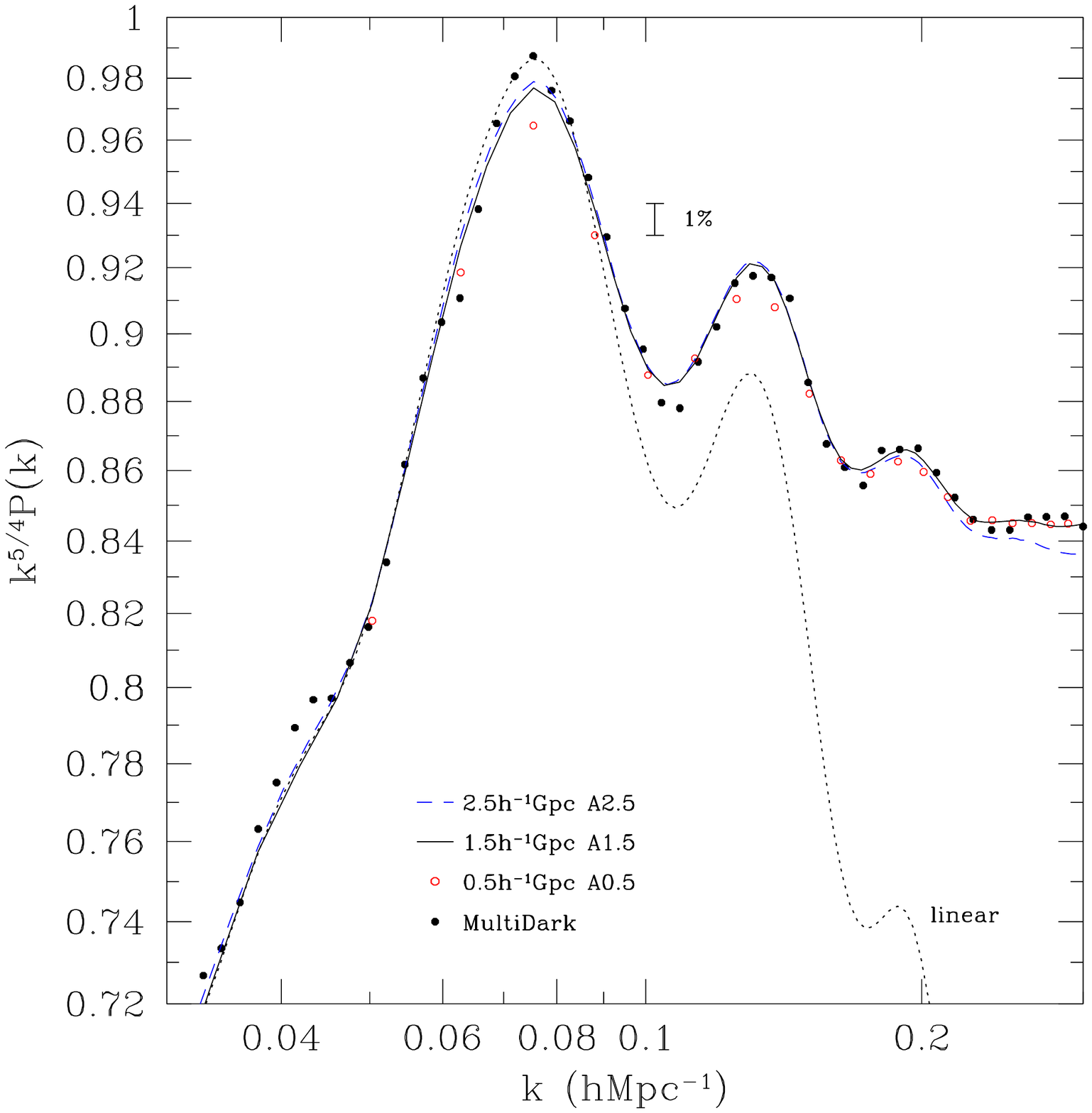}
\caption{Power spectra of dark matter at redshift zero. The linear power
  spectrum is shown as a dotted line. {\it Left:}
  The PPM-GLAM simulations used for the plot
  are A0.5 (red full curve), A1.5 (black dot-dashed) and A2.5 (blue
  dashed). They closely reproduce the clustering of the high-resolution
  MultiDark simulations (black full curve) up to $k\approx 1\,h{\rm Mpc}^{-1}$ with exact
  deviations on larger $k$ depending on the force resolution. 
  {\it Right:} Zoom-in on the region of the BAO peaks. The power spectra were
  multiplied by $k^{5/4}$ and arbitrarily normalised to reveal more
  clearly the differences between the simulations. Because there are only two
  realisations of the MultiDark simulations (BigMDPL \& HMDPL), statistical deviations
  due to cosmic variance are seen at different $k$ (e.g.,
  $k\approx 0.045, 0.07\,h{\rm Mpc}^{-1}$). Cosmic variance of the
  simulations (errors of the mean) are nearly negligible because there
  are thousands of PPM-GLAM realisations. The deviations seen for the A0.5 points (red
  circles) at small $k$ are due to the large bin width, in $k$-space, for this simulation.
}
\label{fig:Converge}
\end{figure*}

The number of time-steps $\Ns$ is proportional to the computational
costs of the simulations. This is why reducing the number of steps is
important for producing a large set of realisations. \citet{QPM} and
\citet{Koda2016} use just $\sim 10$ time-steps for their QPM
and COLA simulations. \citet{Feng2016} and \citet{Izard2015} advocate using
$\Ns\approx 40$ steps for Fast-PM and ICE-COLA. The question still
remains: what optimal number of time-steps should be adopted?  However, there
is no answer to this question without specifying the required force resolution,
and without specifying how the simulations will be used to generate
mock galaxies.

In Appendix~\ref{sec:AppendixB} we provide a detailed discussion on
the effects of time-stepping. We argue that for the stability and
accuracy of the integration of the dark matter particle trajectories
inside dense (quasi-) virialised objects, such as clusters of
galaxies, the time-step $\Delta t$ must be smaller enough to satisfy
the constraints given by eqs.~(\ref{eq:dtcondition}) and
(\ref{eq:Stability}). For example, FastPM simulations with 40
time-steps and force resolution of $\Delta x = 0.2\,\Mpch$\
\citep[see][]{Feng2016} do not satisfy these conditions and would
require 2-2.5 times more time-steps. However, a small number of
time-steps manifests itself not in the power spectrum (though, some
decline in $P(k)$ happens at $k\sim 1\,\Mpch$). Its effect is mostly
observed in a significantly reduced fraction of volume with large
overdensities and random velocities, which potentially introduces
undesirable scale-dependent bias.

Because our main goal is to produce simulations with the minimum
corrections to the local density and peculiar velocities, we use
$\Ns\approx 100-200$ time-steps in our PPM-GLAM simulations. This
number of steps also removes the need to split particle displacements
into quasi-linear ones and the deviations from quasi-linear
predictions. Thus, in this way we greatly reduce the complexity of the
code and increase its speed, while also substantially reduce the memory
requirements.

\makeatletter{}\section{Simulations}
\label{sec:sim}

We made a large number of PPM-GLAM simulations -- about 15,000 -- to
study different aspects of the clustering statistics of the dark
matter density field in the flat $\LCDM$ Planck cosmology. The
numerical parameters of our simulations are presented in
Table~\ref{table:simtable}. All the simulations were started at
initial redshift $z_{\rm init}=100$ using the Zeldovich approximation.
The simulations span three orders of magnitude in mass resolution, a factor of
six in force resolution and differ by a factor of 500 in effective
volume. To our knowledge this is the largest set of simulations
available today. Power spectra and covariance matrices are publicly
available in our $\textit{Skies and Universes}$
site\footnote{http://projects.ift.uam-csic.es/skies-universes/}.

The PPM-GLAM simulations labeled with letter A are the main simulations used for
estimates of the power spectrum and the covariance matrix. Series B and C
are designed to study different numerical effects. In particular, C1a
are actually four simulations run with the different number of
steps: $\Ns =34, 68, 147, 302$. There are also four C1b simulations
that differ by the force resolution: $\Ng =1000, 2000, 3000, 4000$.

We compare the performance of our PPM-GLAM simulations with the
results obtained from some of the MultiDark 
simulations\footnote{http://www.multidark.org} run with L-Gadget2:
BigMDPL and HMDMPL \citep[see for details][]{Klypin2016}. The
parameters of these large-box and high-resolution simulations are also
presented in Table~\ref{table:simtable}. For comparison we also list
very large number of low resolution simulations performed by
\citet{Takahashi2009}, \citet{Li2014} and \citet{Blot2015} with the
Gadget2, L-Gadget2 and AMR codes, respectively, to study the power
spectrum covariances. Details of the QPM \citep{QPM} and COLA
\citep{Koda2016} simulations that were used to generate a large number
of galaxy mocks for the BOSS and WiggleZ galaxy redshift surveys are also
given in Table~\ref{table:simtable}. Note that the QPM simulations
have very low force resolution, which requires substantial modeling
 on how dark matter should be clustered and moving on the scale of
galaxies.

We estimate the power spectrum $P(k)$ of the dark matter density field
in all our 10 simulation snapshots for each realisation, but in this
paper we mostly focus on the $z=0$ results. For each simulation we
estimate the density on a 3D-mesh of the size indicated in
Table~\ref{table:simtable}. We then apply FFT to generate the
amplitudes of the Fourier harmonics $\delta_{i,j,k}$ in
phase-space. The spacing of the Fourier harmonics is equal to the
length of the fundamental harmonic $\Delta\kappa \equiv 2\pi/L$. Thus,
the wave-vector ${\bf k}_{i,j,k}$ corresponding to each triplet
$(i,j,k)$, where $i,j,k = 0,\dots \Ng-1$, is
${\bf k}_{i,j,k}= \left(i\Delta\kappa,
  j\Delta\kappa,k\Delta\kappa\right)$.
Just as the spacing $\Delta x = L/\Ng$ in real-space represents the
minimum resolved scale (see Sec.~\ref{sec:ppmglam}), the fundamental
harmonic $\Delta\kappa$ is the minimum spacing in Fourier-space,
i.e. one cannot probe the power spectrum below that scale. To estimate
the power spectrum we use a constant bin size equal to
$\Delta\kappa$. This binning results in very fine binning at high
frequencies, but preserves the phase resolution at very small
frequencies (long waves).

A correction is applied to the power spectrum to compensate the
effects of the CIC density assignment:
$P(k) = P_{\rm raw}(k)/\left[1 -(2/3)\sin^2(\pi k/2k_{\rm Ny})
\right]$,
where the Nyquist frequency of the grid is
$k_{\rm Ny} =(\Ng/2)\Delta\kappa=\pi/\Delta x$. The same number of
grid points is used for estimates of the power spectrum as for running
the simulations. We typically use results only for
$k< (0.3-0.5)k_{\rm Ny}$.  No corrections are applied for the finite
number of particles because these are small for the scales and
particle number-densities considered in this paper.

Similar to CIC in real space, we apply CIC filtering in  Fourier
space. For each Fourier harmonic ${\bf k}_{i,j,k}$ the code finds left
and right bins by dividing the length of the wave-vector by the
fundamental harmonic $\Delta\kappa$ and then by taking the integer
part ${\rm INT}(|{\bf k}_{i,j,k}|/\Delta\kappa)$. The contributions to the
left and right bins are found proportionally to the difference between
harmonic and  bin wave-numbers. This procedure reduces the noise in
the power spectrum by $\sim 30\%$ at the cost of introducing
dependencies in power spectrum values in adjacent bins in  Fourier
space. Effects of this filtering are included in estimates of the
covariance matrix: they mostly change (reduce) diagonal components.

\makeatletter{}\begin{figure*}
\centering
\includegraphics[width=0.49\textwidth]
{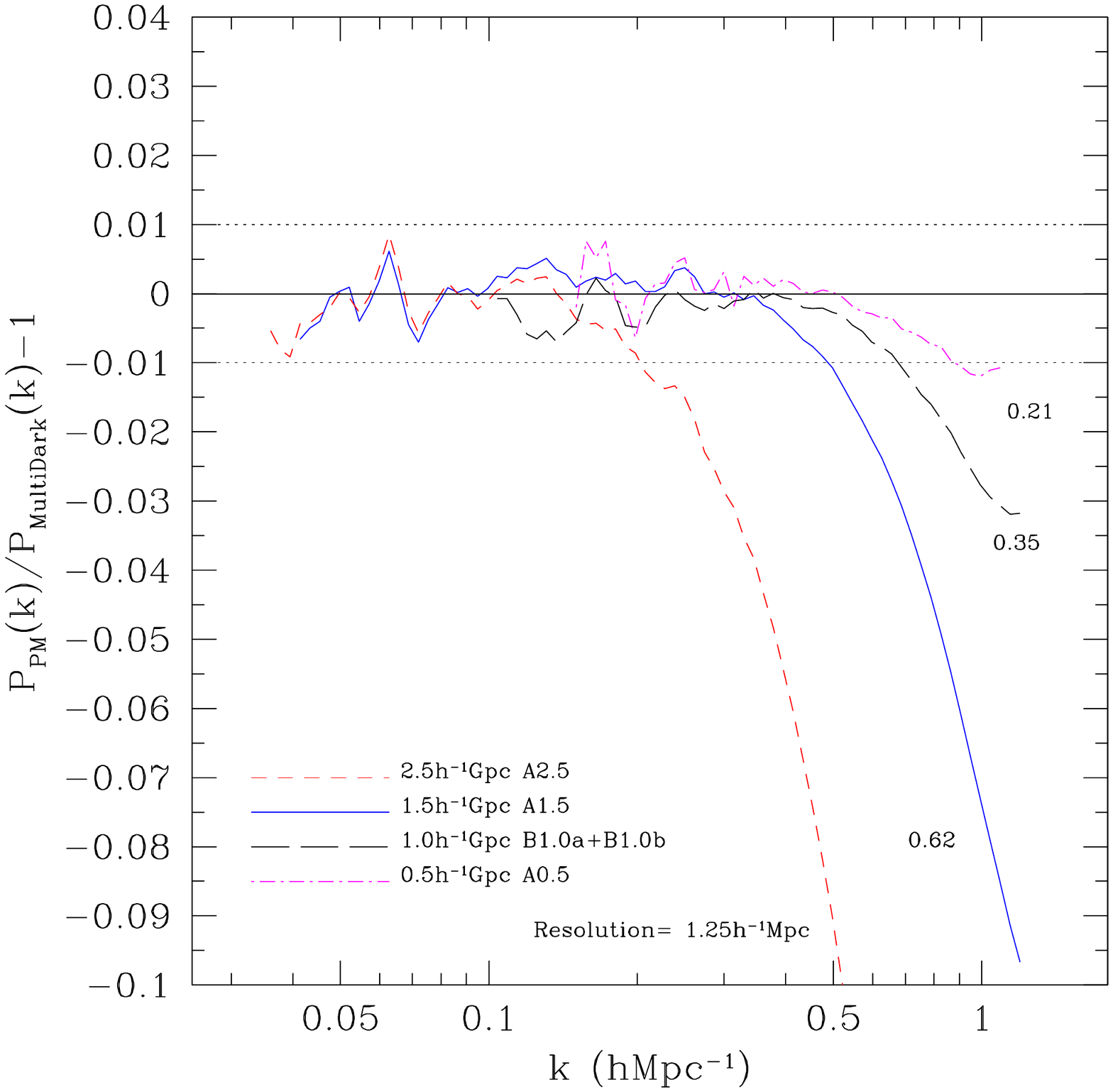}
\includegraphics[width=0.49\textwidth]
{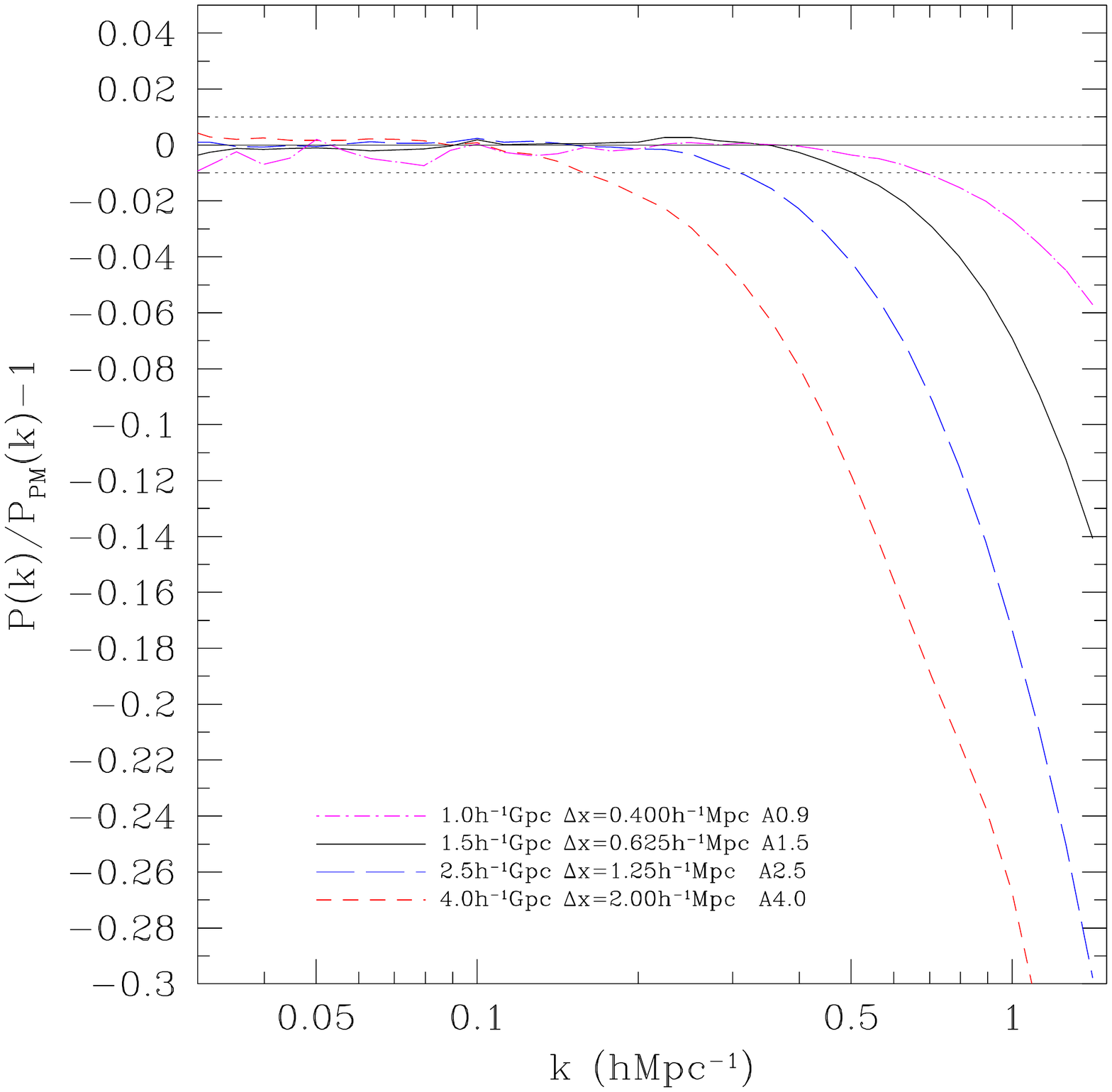}
\caption{Convergence of power spectra in real-space at redshift
  zero. {\it Left:} Ratios of the real-space power spectra in PPM-GLAM
  simulations to the power spectrum of the MultiDark simulations.
  Simulations used for the plot are indicated in the plot. For the
  black long dashed curve we use the average of B1.0a and B1.0b
  simulations.  The lack of force resolution results in the decline of the
  power spectrum at large wave-numbers. As the resolution increases,
  the power spectrum becomes closer to the MultiDark results. At low
  values of $k \lsim 0.3\,h{\rm Mpc}^{-1}$ deviations at
  $\sim 0.5\%$ level are caused by cosmic variance in the MultiDark
  simulations. Otherwise there are no systematics related with the
  finite box size.  
  {\it Right:} Convergence of power spectra in PPM-GLAM
  simulations with different box sizes and resolutions. We plot the ratios
  of power spectra of A0.9, A1.5, A2.5 and A4.0 simulations to the combined power
  spectrum $P_{\rm PM}$ that is found by averaging best simulations
  for different ranges of wave-numbers (see text). Here the effects of cosmic
  variance (seen on the left panel) are negligible because of the
  averaging over thousands of realisations. Missing waves longer than
  the simulation boxes have little effect for $L\gsim 1\,\Mpch$ and
  $k>0.05\,h{\rm Mpc}^{-1}$.}
\label{fig:ConvergePk}
\end{figure*}

\makeatletter{}\section{Power Spectrum}
\label{sec:powersp}

We study the convergence performance of the power spectra obtained
with PPM-GLAM by comparing our results with those drawn from the
high-resolution MultiDark simulations listed in
Table~\ref{table:simtable}. Specifically, we average power spectra of
BigMDPL and HMDPL simulations weighted proportianally to the volume of each
simulation. Left panel in Figure~\ref{fig:Converge} shows the power
spectra in $log-log$ scale for the A0.5 (red full curve), A1.5 (black
dot-dashed) and A2.5 (blue dashed) PPM-GLAM simulations. As a
reference we also show the linear power spectrum (dotted line).

Our simulations closely reproduce well the clustering of the
high-resolution MultiDark simulations (black full curve) both for
long-waves (as may have been expected) and even for larger
wave-numbers up to $k\approx 1\,h{\rm Mpc}^{-1}$ with exact deviations
on larger $k$ depending on force resolution. The lack of force
resolution results in the decline of the power spectrum at large
wave-numbers: as resolution increases the power spectrum becomes
closer to the MultiDark results.
This is clearly seen in the left panel of Figure~\ref{fig:ConvergePk}
where we show the ratios of the PPM power spectra $P(k)$ of A0.5 (red
dot-dashed curve), A1.5 (blue full curve) and A2.5 (red dashed curve)
to the power spectrum obtained from the MultiDark simulations. We also
label in the plot the force resolution for each of the PPM-GLAM
simulations. The results for the A0.5 simulations are presented only
for $k>0.15\,h{\rm Mpc}^{-1}$ because the bin smearing becomes visible
($\sim 2$\%) at lower frequencies due to the small volume of each
individual realisation. We also plot the average of the B1.0a and
B1.0b simulations (black long-dashed curve).  Again, the deviations
are less than 1\% on large scales and they start to increase as we go
to larger $k$ with the magnitude of the error depending on the force
resolution. Note that the ratios of the PPM-GLAM results to those in
the MultiDark simulations are the same at long-waves with
$k<0.1\,h{\rm Mpc}^{-1}$. This is related with the cosmic variance present in
the MultiDark $P(k)$ since there are only two realisations,
i.e. BigMDPL and HMDPL.

The right panel in Figure~\ref{fig:Converge} zooms-in on the relevant
domain $k\approx 0.07-0.2\,h{\rm Mpc}^{-1}$ of the BAO peaks. In this
plot the power spectrum $P(k)$ is multiplied by the factor $k^{5/4}$
to reduce the dynamical scale allowing us to see the differences as
small as a fraction of percent. The cosmic variance of the PPM-GLAM
simulations (errors of the mean) are nearly negligible because there
are thousands of realisations. The observed deviations of the A0.5
points at small $k$ (e.g.  $k\approx 0.08\,h{\rm Mpc}^{-1}$) are due
to the large size of the binning in $k$ space defined by the width of
the fundamental harmonic $\Delta k=2\pi/L=0.0125\,h{\rm Mpc}^{-1}$. If
we consider simulations with small binning, i.e. simulation box
$L\gsim 1\,\Gpch$, then the deviations from the MultiDark simulations
are less than 1~per cent on the large scales.

\makeatletter{}\begin{figure*}
\centering
\includegraphics[width=0.49\textwidth]
{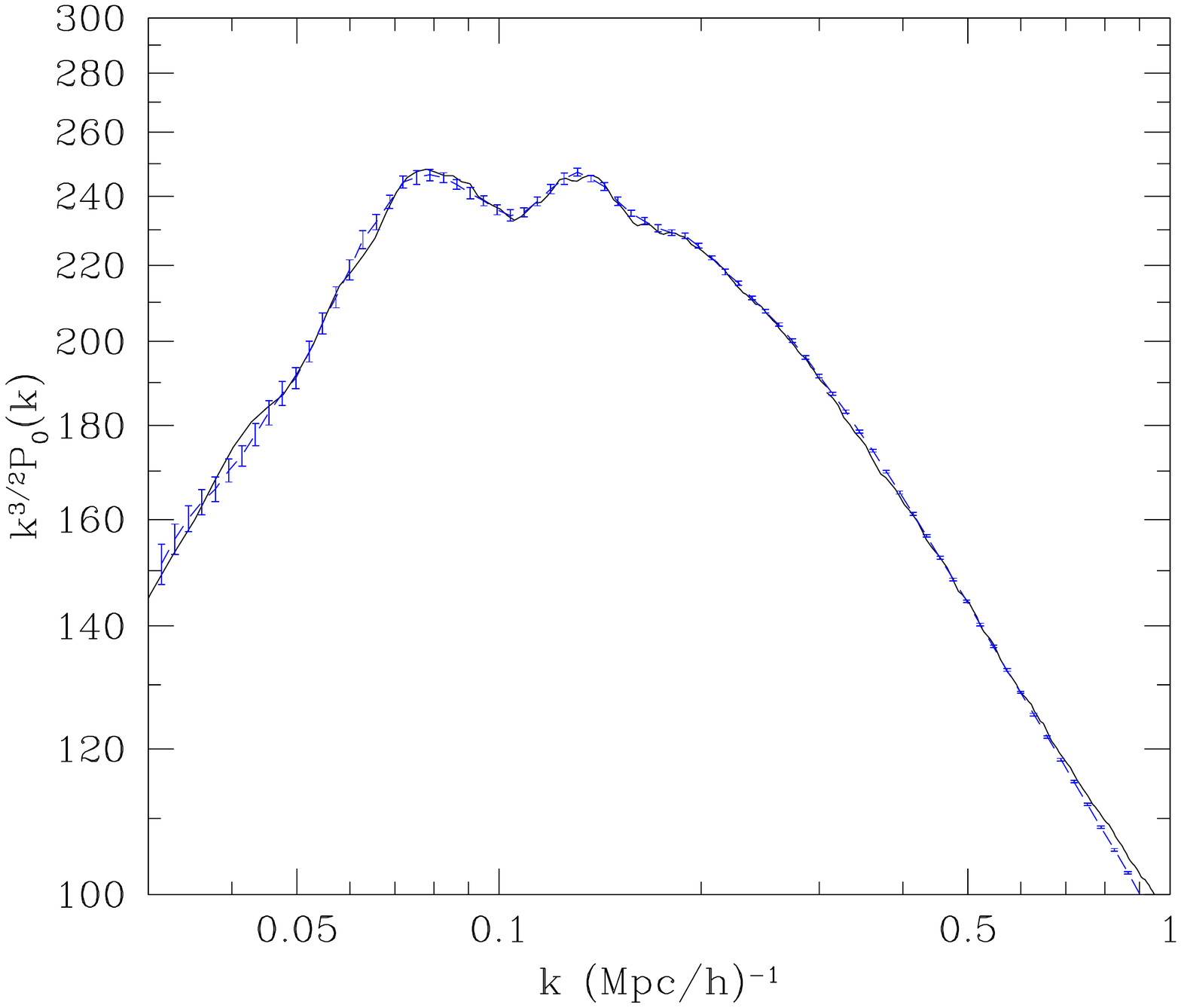}
\includegraphics[width=0.49\textwidth]
{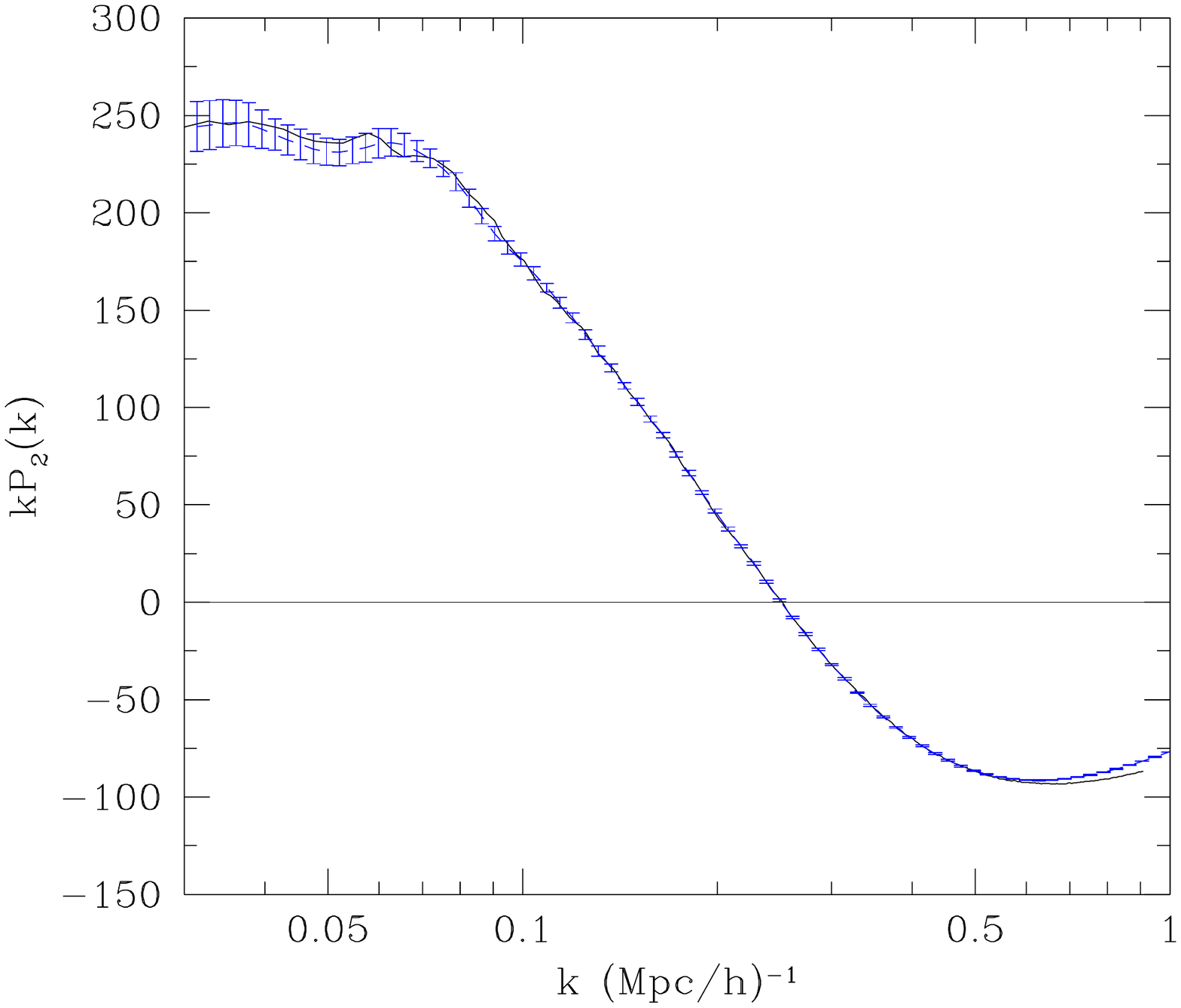}
\caption{Comparison of power spectra in redshift-space at redshift
  zero. Monopole (left panel) and quadrupole (right panel) were scaled
  with different power of $k$ to reduce the dynamical range of the
  plots. The solid black curves correspond to the BigMDPL simulation
  estimated using 5\% of all particles. The dashed curves and error
  bars are for the B1.5 simulations.}
\label{fig:ConvergeZ}
\end{figure*}

This is clearly seen in the left panel of Figure~\ref{fig:ConvergePk}
were we show the ratios of GLAM power spectra to $P(k)$ in the
MultiDark simulations. Results for the A0.5 simulations are presented
only for $k>0.15\,h{\rm Mpc}^{-1}$ because the bin smearing becomes
visible ($\sim 2$\%) at lower frequencies. Again, the deviations are
less than 1\% on large scales and start to increase as we go to large
$k$ with the magnitude of the error depending on the force
resolution. Note that the ratios of the PM results to those in the
MultiDark simulations are the same at $k<0.1\,h{\rm Mpc}^{-1}$. This
is related with the cosmic variance in the MultiDark $P(k)$ -- there
are only two realizations of MultiDark.

In order to test the effects of force resolution $\epsilon$ and finite
box size $L$ we construct the combined power spectrum by taking the
average of the best PPM-GLAM simulations in separate ranges of
frequency: (1) for $k<0.1\,h{\rm Mpc}^{-1}$ we average $P(k)$ of all
A1.5, A2.5, and A4.0 realisations, (2) for the range
$0.1\,h{\rm Mpc}^{-1}< k < 0.2\,h{\rm Mpc}^{-1}$ we take the average
of the A2.5, A1.5, A0.9 simulations, (3) for
$0.2\,h{\rm Mpc}^{-1}< k < 0.4\,h{\rm Mpc}^{-1}$, A0.9 and A0.5
simulations are used, and (4) for larger wave-numbers we consider only
the A0.5 realisations. We show in Figure~\ref{fig:ConvergePk} (right
panel) the ratios of $P(k)$ of A0.9, A1.5, A2.5 and A4.0 simulations
to the combined power spectrum. The deviations of each simulation from
this combined $P_{\rm PM}(k)$ spectrum is a measure of the errors in
each simulation. Note that these errors are broadly consistent with
those of MultiDark except for the very long waves where we now do not
have artificial deviations due to the cosmic variance. This plot gives
us means to estimate what resolution is needed if we are required to
provide some specified accuracy at a given scale. For example, if the
errors in $P(k)$ should be less than 1\% at $k<0.5\,h{\rm Mpc}^{-1}$,
then the resolution of the simulation should be $\Delta x
=0.62\Mpch$.
For, 1\% at $k<1\,h{\rm Mpc}^{-1}$ the resolution should be
$\Delta x =0.2\Mpch$.

The right panel in Figure~\ref{fig:ConvergePk} also gives very useful
information on the effects due to the finite box size. The size of the
computational box is an important factor that affects the total CPU
time, the statistics of the large-scale fluctuations, and possibly the
non-linear coupling of long- and short waves. Non-linear coupling
effects are of some concern
\citep[e.g.,][]{Gnedin2011,Li2014,Wagner2015,Baldauf2016} because the
long-waves missed in small-box simulations (called Super Sample
Covariance or SSC) can affect both the power spectrum and the
covariance matrix. The magnitude of the SSC effect critically depends
on the size of the computational box. Because our main target is
relatively large boxes $L\approx (1-1.5)\,\Mpch$ with high resolution,
missing waves longer than these scales are deeply in the linear
regime, and thus the SSC effects are expected to be small, which the
right panel in Figure~\ref{fig:ConvergePk} clearly demonstrates.

The SSC effects should manifest themselves as an increase in $P(k)$ at
small $k$ in simulations with large $L$ as compared with simulations
with smaller $L$. Indeed, we see this effect at very long-waves. For
example, at $k=0.03\,h{\rm Mpc}^{-1}$ the power spectrum in A0.5
simulations was below that of A4.0 by 4\%. However, the effect
becomes much smaller as the box size increases. The error becomes less than
0.2\% for the A1.5 simulations. It is also much smaller for shorter
waves. For example, the error is less than 0.5\% for A0.9 simulations at
$k>0.05\,h{\rm Mpc}^{-1}$. This can be understood if one estimates the
amplitude of density fluctuations in waves longer than the simulation
box. For $L=500\,\Mpch$ and for the Planck cosmology the $rms$ density
fluctuation $\delta(>L)$ is relatively large:
$\delta(>500\,\Mpch)=0.027$. It is nearly ten times smaller for A1.5
simulations: $\delta(>1500\,\Mpch)=0.0036$. 

While the main interest of this paper is in the clustering in
real-space, we also tested dark matter clustering in redshift-space
and compared that with the MultiDark simulations. We plot in
Figure~\ref{fig:ConvergeZ} both monopole (left panel) and quadrupole
(right panel) for PPM-GLAM B1.5 and BigMDPL, which shows a remarkable
agreement.

\makeatletter{}\section{Covariance Matrix.}
\label{sec:covmatrix}
\makeatletter{}\begin{figure*}
\centering
\includegraphics[width=0.495\textwidth]
{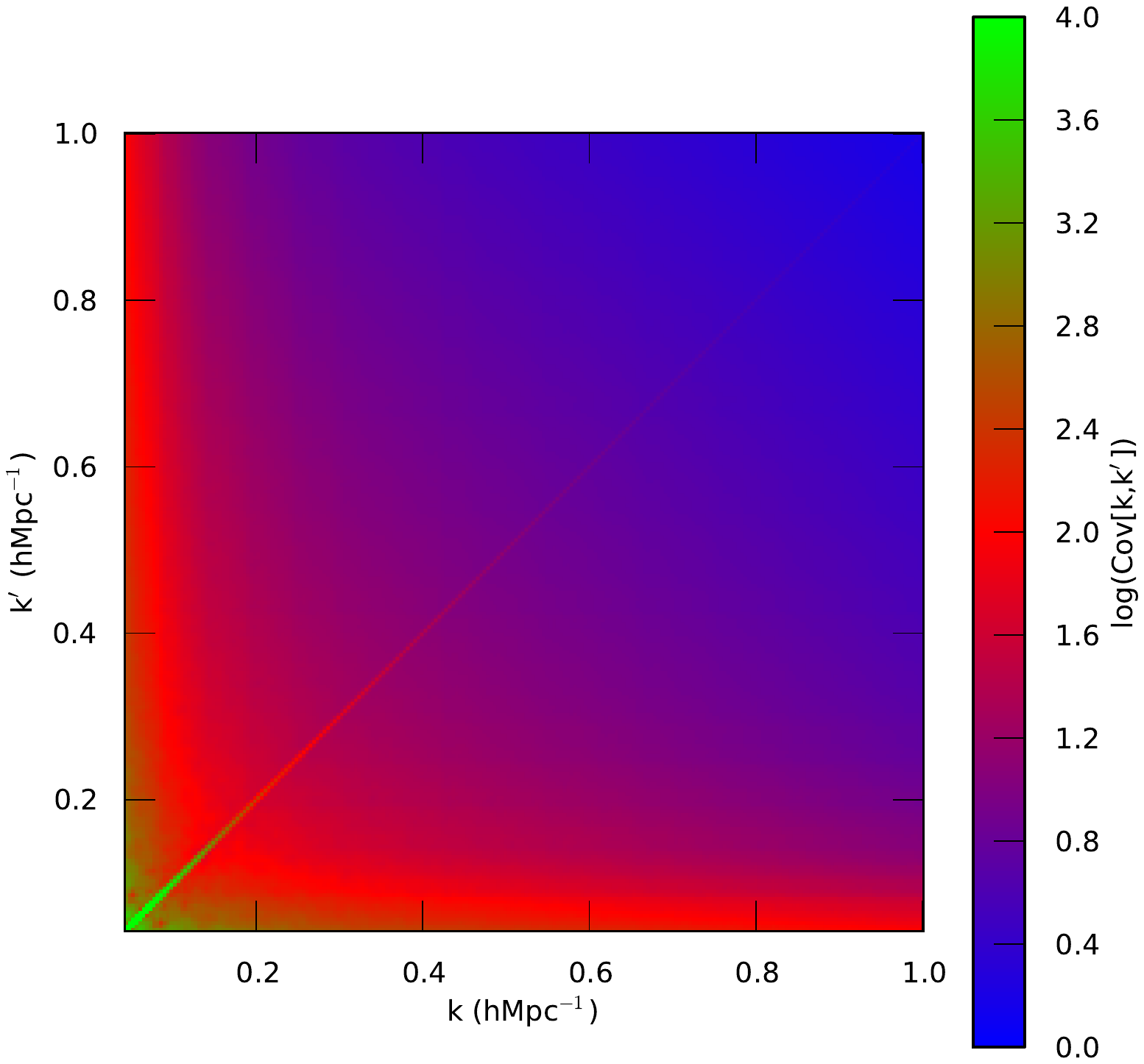}
\includegraphics[width=0.495\textwidth]
{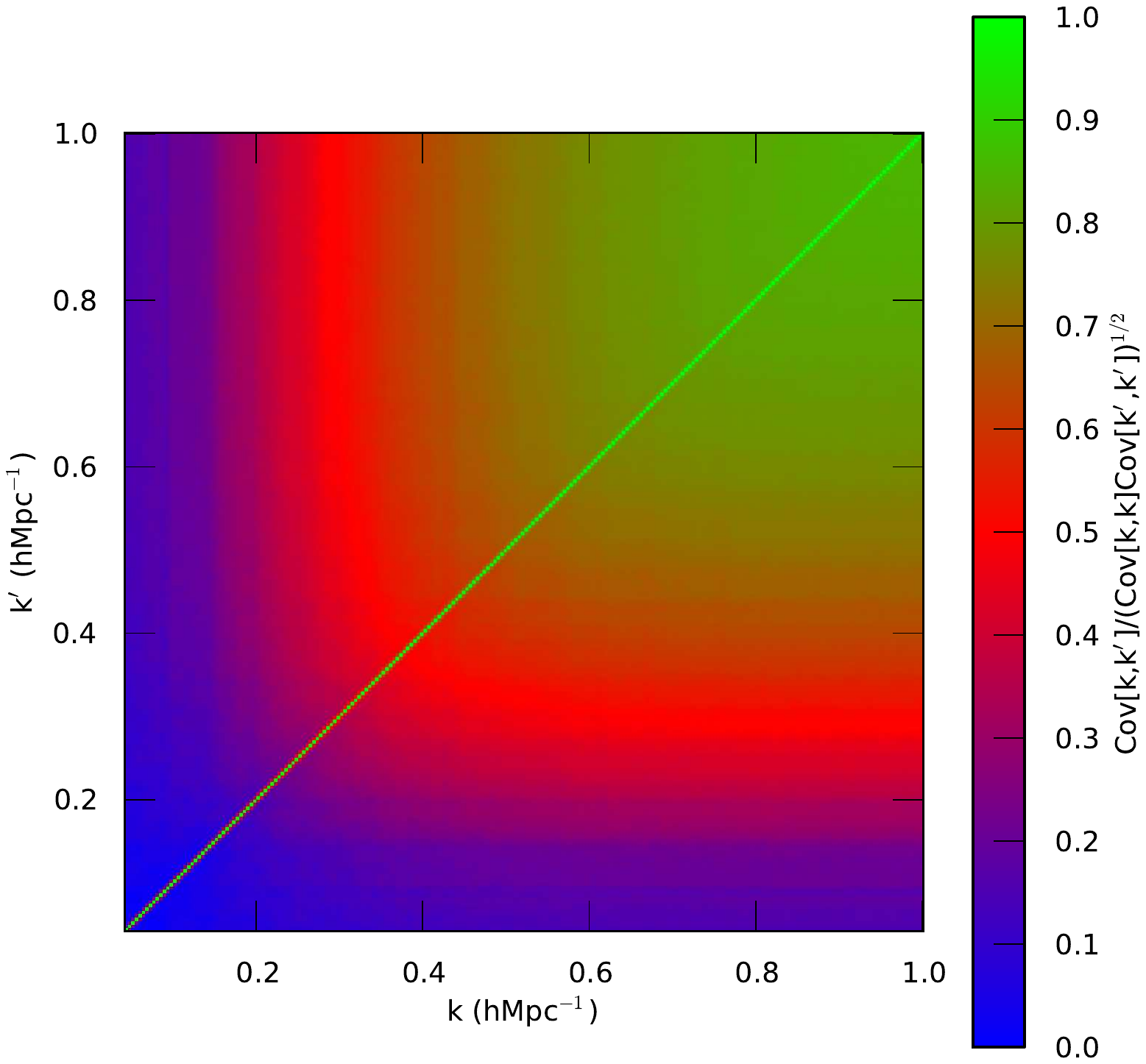}
\caption{Covariance matrix $C(k,k\prime)$ on logarithmic
  scale ({\it left panel}) and covariance coefficient
  $C(k,k\prime)/\sqrt{C(k,k)C(k\prime,k\prime)}$ ({\it right panel}) of the dark matter
  power spectrum at redshift zero for the PPM-GLAM A1.5 simulations. With the
  exception of the narrow spike at the diagonal of the matrix, the
  covariance matrix is a smooth function. Horizontal and vertical
  stripes seen in the correlation coefficient at small $k$ are due to
  cosmic variance.}
\label{fig:CovMatrix}
\end{figure*}

The covariance matrix $C(k,k^\prime)$ is defined as a reduced cross
product of the power spectra at different wave-numbers for the same
realisation averaged over different realisations, i.e.
\begin{equation}
  C(k,k^\prime) \equiv  \langle P(k)P(k^\prime)\rangle - 
                  \langle P(k)\rangle\langle P(k^\prime)\rangle.
\label{eq:Cov}
\end{equation}
The diagonal and non-diagonal components of the covariance matrix have
typically very different magnitudes and evolve differently with
redshift. Their diagonal terms are larger than the off-diagonal ones,
but there are many more off-diagonal terms making them cumulatively
important \citep{Taylor2013,Percival2014,OConnell2016}. Off-diagonal
terms are solely due to non-linear clustering effects: in a
statistical sense the off-diagonal terms are equal to zero in the
linear regime. The diagonal component $C(k,k)$ can be written as a
sum of the gaussian fluctuations due to the finite number of harmonics
in a bin and terms that are due to non-linear growth of fluctuations:
\begin{equation}
C(k,k) \equiv C_{\rm Gauss}(k) + C_{\rm non}(k,k),
\end{equation}
where the Gaussian term depend on the amplitude of the power spectrum
$P(k)$ and on the number of harmonics $N_h$:
\begin{equation}
C_{\rm Gauss}(k) = \alpha\frac{2}{N_h}P^2(k),\quad N_h=\frac{4\pi k^2\Delta k}
                                                      {\left(2\pi/L\right)^3},
\label{eq:Gauss}
\end{equation}
where the numerical factor $\alpha$ is equal to unity for the
Nearest-Grid-Point (NGP) assignment in Fourier-space and
$\alpha=2/3$ for the CIC assignment used in this paper.
Note that for a fixed bin width $\Delta k$ the number of harmonics,
and thus, the amplitude of the Gaussian term scales proportional to the
computational volume $N_h \propto L^3$ with $C_{\rm Gauss}(k)\propto 1/L^3$.
 
There are two ways of displaying the covariance matrix. One can
normalise $C(k,k^\prime)$ by its diagonal component:
$r(k,k^\prime) \equiv
C(k,k^\prime)/\sqrt{C(k,k)C(k^\prime,k^\prime)}$.
This quantity is called the correlation coefficient, and by definition,
$r(k,k)\equiv 1$. The covariance matrix can also be normalised by the
"signal", i.e. the product of power spectra at the two involved
wave-numbers: $\sqrt{C(k,k^\prime)/P(k)P(k^\prime)}$.
\makeatletter{}\begin{figure*}
\centering
\includegraphics[width=0.49\textwidth]
{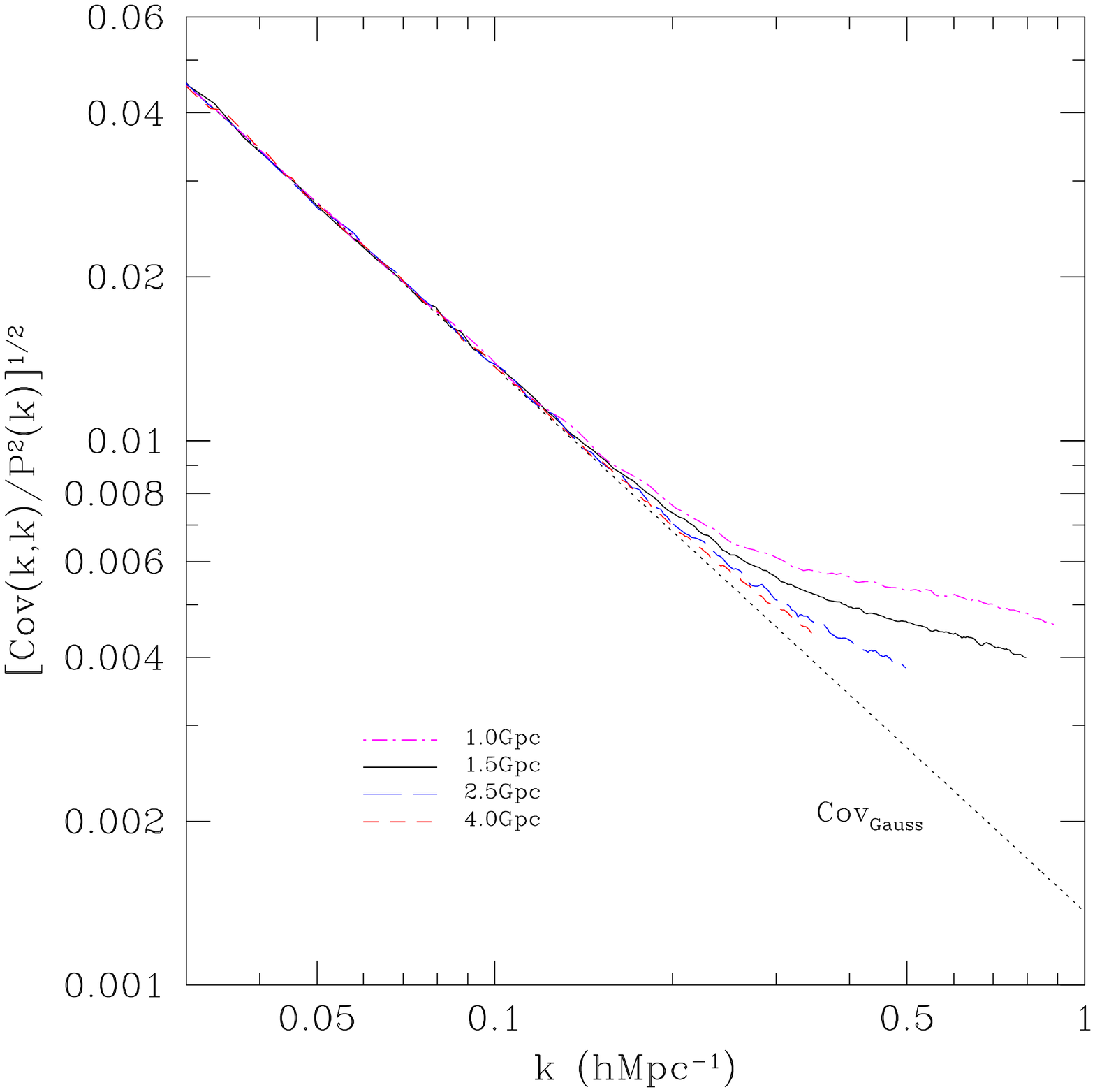}
\includegraphics[width=0.49\textwidth]
{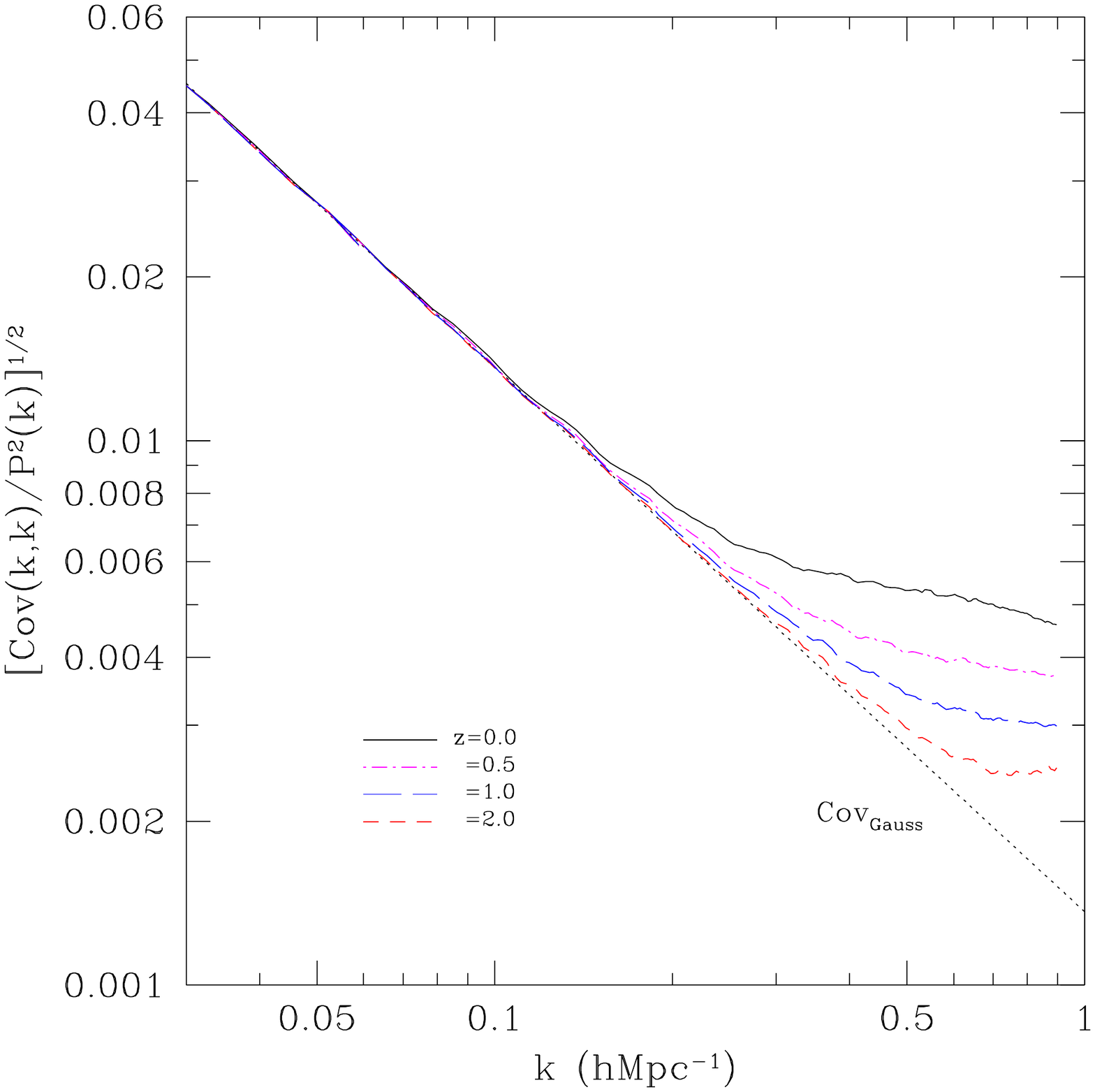}
\caption{Diagonal components of the covariance matrix for PPM-GLAM simulations
  with different box sizes at $z=0$ (left panel), and for different
  redshifts for the A0.9 simulations (right panel). All results were
  rescaled to the simulation volume of $1.5\Gpch$. The dotted lines show
  the Gaussian contribution $C_{\rm Gauss}$ given by
  eq.~(\ref{eq:Gauss}) with $\alpha=2/3$, which gives a good
  approximation up to $k\lsim 0.2\,h{\rm Mpc}^{-1}$. At larger
  wave-numbers the covariance matrix is substantially larger than the
  Gaussian term due to the non-linear coupling of waves. The
  diagonal terms evolve with redshift and sensitively depend on
  the force resolution.}
\label{fig:Diagonal}
\end{figure*}

\makeatletter{}\begin{figure*}
\centering
\includegraphics[width=0.49\textwidth]
{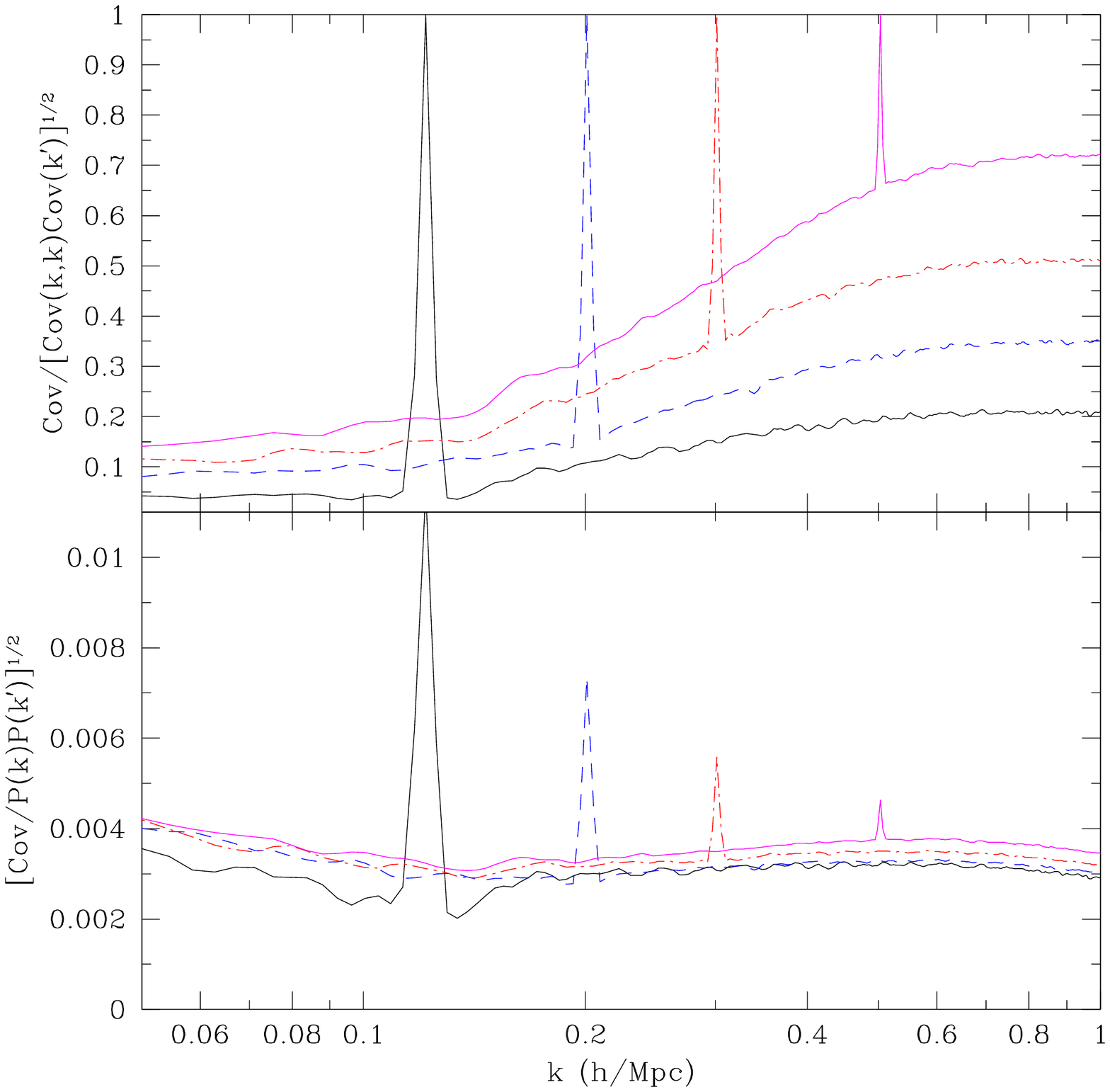}
\includegraphics[width=0.49\textwidth]
{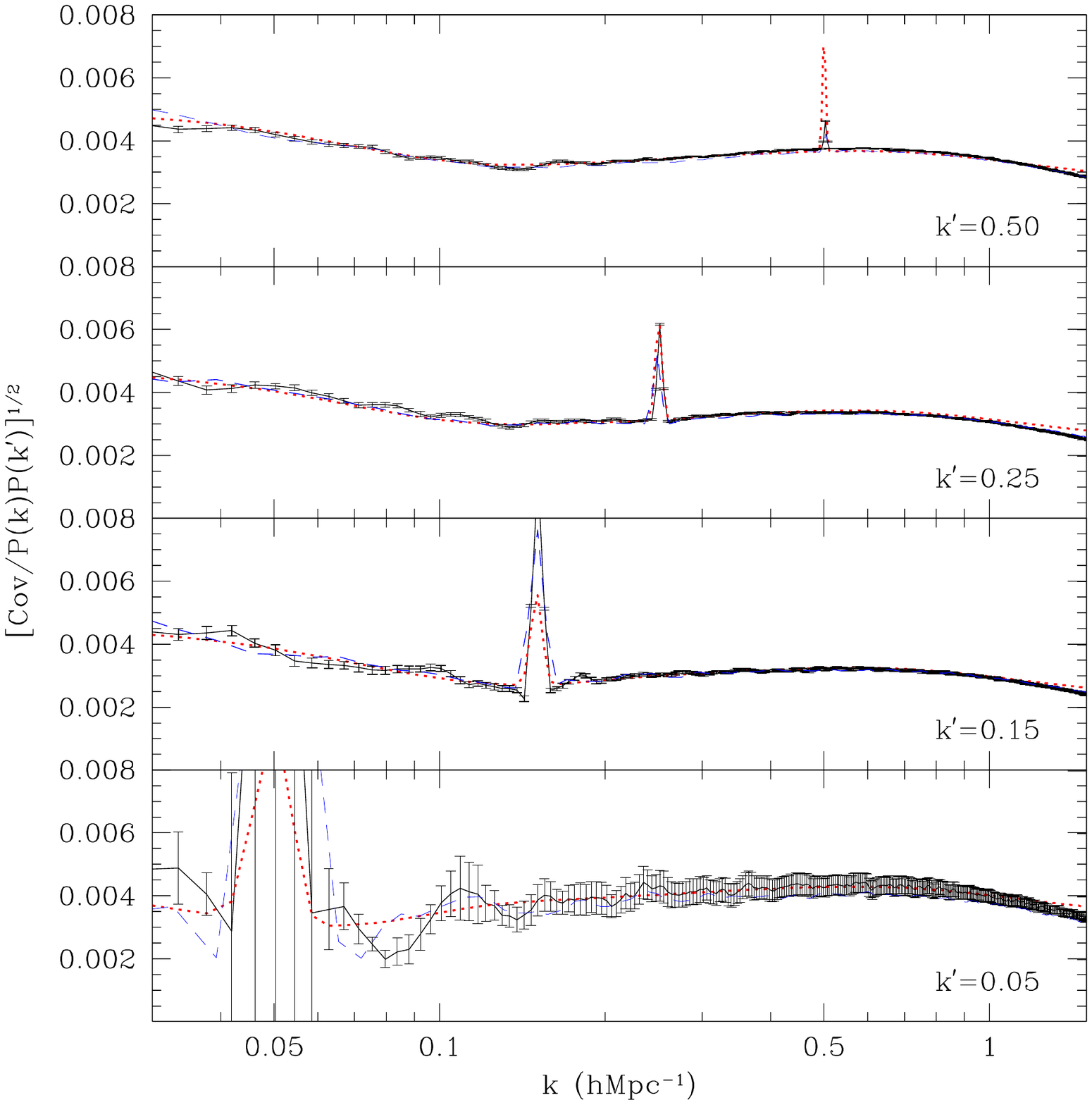}
\caption{Slices through the $z=0$ covariance matrix $C(k,k\prime)$ at
  different values of $k^\prime$ for $\sim 4500$ realisations of
  PPM-GLAM A1.5.  {\it Left:} Covariance coefficient
  $C(k,k\prime)/[C(k,k)C(k\prime,k\prime)]^{1/2}$ (top panel) and
  covariance matrix normalised to the power spectrum
  $[C(k,k\prime)/P(k)P(k^\prime)]^{1/2}$ (bottom panel) for
  $k^\prime = 0.12, 0.2, 0.3, 0.5\,{\rm h\,{\rm Mpc}}^{-1}$.  Note the
  change in scale of the y-axes. The covariance matrix normalised by
  the signal is very small $\sim 3\times 10^{-3}$ and relatively flat
  as compared with the large differences seen in the covariance
  coefficients on the top panel. {\it Right:} Detailed views of the
  covariance matrix. Solid curves with the error bars are for the A1.5
  simulations. Results for the A0.9 simulations scaled to the
  1.5\Gpch\ volumes are shown with dashed curves.  Dotted curves show
  the analytical approximation given in Eqs.~(11-13). As $k$ increases
  the covariance matrix first decreases, reaches the minimum at
  $k\approx (0.1-0.2)h\,{\rm Mpc}^{-1}$ and then has a maximum at
  $k\approx (0.5-0.6)h\,{\rm Mpc}^{-1}$. In addition, it has a dip on
  both sides of the diagonal components.}
\label{fig:CovZ0}
\end{figure*}

Figure~\ref{fig:CovMatrix} shows the covariance matrix $C(k,k^\prime)$ and the correlation
coefficient for the PPM-GLAM A1.5 simulations at $z=0$. With the exception of a
narrow spike at the diagonal of the matrix, the covariance matrix is a
smooth function. Horizontal and vertical stripes seen in the
correlation coefficient at small $k$ are due to cosmic
variance. They gradually become weaker as the number of realisations
increases \citep[e.g.,][]{Blot2015}.

The diagonal terms of the covariance matrix are presented in
Figure~\ref{fig:Diagonal}. In the left panel we compare the results of
various simulations at $z=0$ with different box sizes. In order to do
that, we rescale the individual $C(k,k)$ to that of the volume of the
A1.5 simulation with $(1.5\Gpch)^3$.  Up to
$k\lsim 0.2\,h{\rm Mpc}^{-1}$ the covariance matrix is well described
by the Gaussian term, but at larger wave-numbers it substantially
exceeds the Gaussian contribution due to the non-linear coupling of
waves. The force resolution plays an important role here.

There are no indications that SSC waves (modes longer than the
simulation box) affect the diagonal components. If present, SSC
effects should results in enhanced $C(k,k)$ in simulations with very
large simulation boxes \citep{Li2014,Baldauf2016}. For
example, A4.0 results should have a larger covariance matrix as
compared with that of the A0.9 simulations. However, the right panel of
Figure~\ref{fig:Diagonal} clearly shows the opposite effect: A0.9
results are {\it above} A4.0  presumably due to the better force resolution
that produces larger non-linear effects.

\makeatletter{}\begin{figure*}
\centering
\includegraphics[width=0.47\textwidth]
{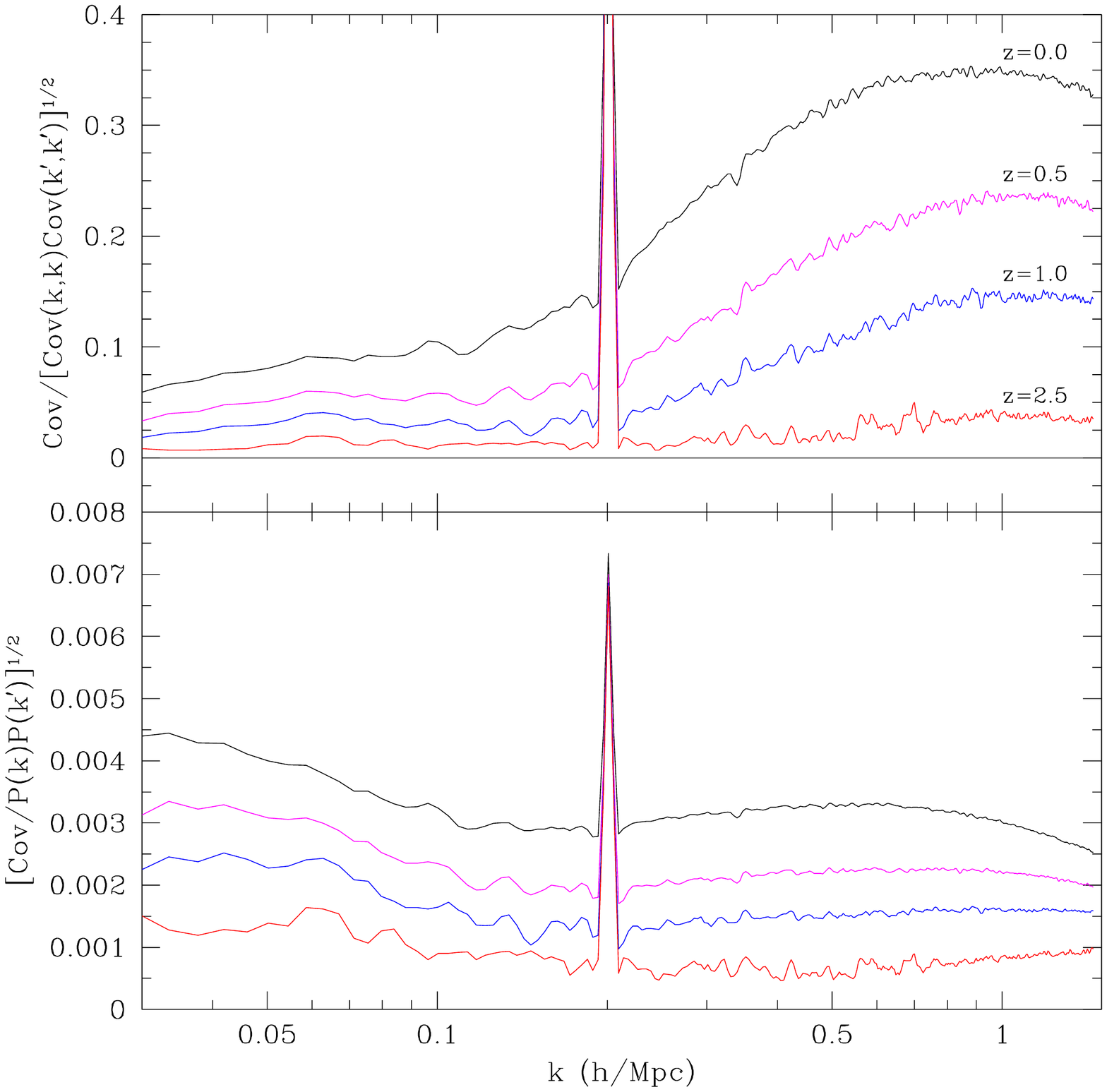}
\includegraphics[width=0.47\textwidth]
{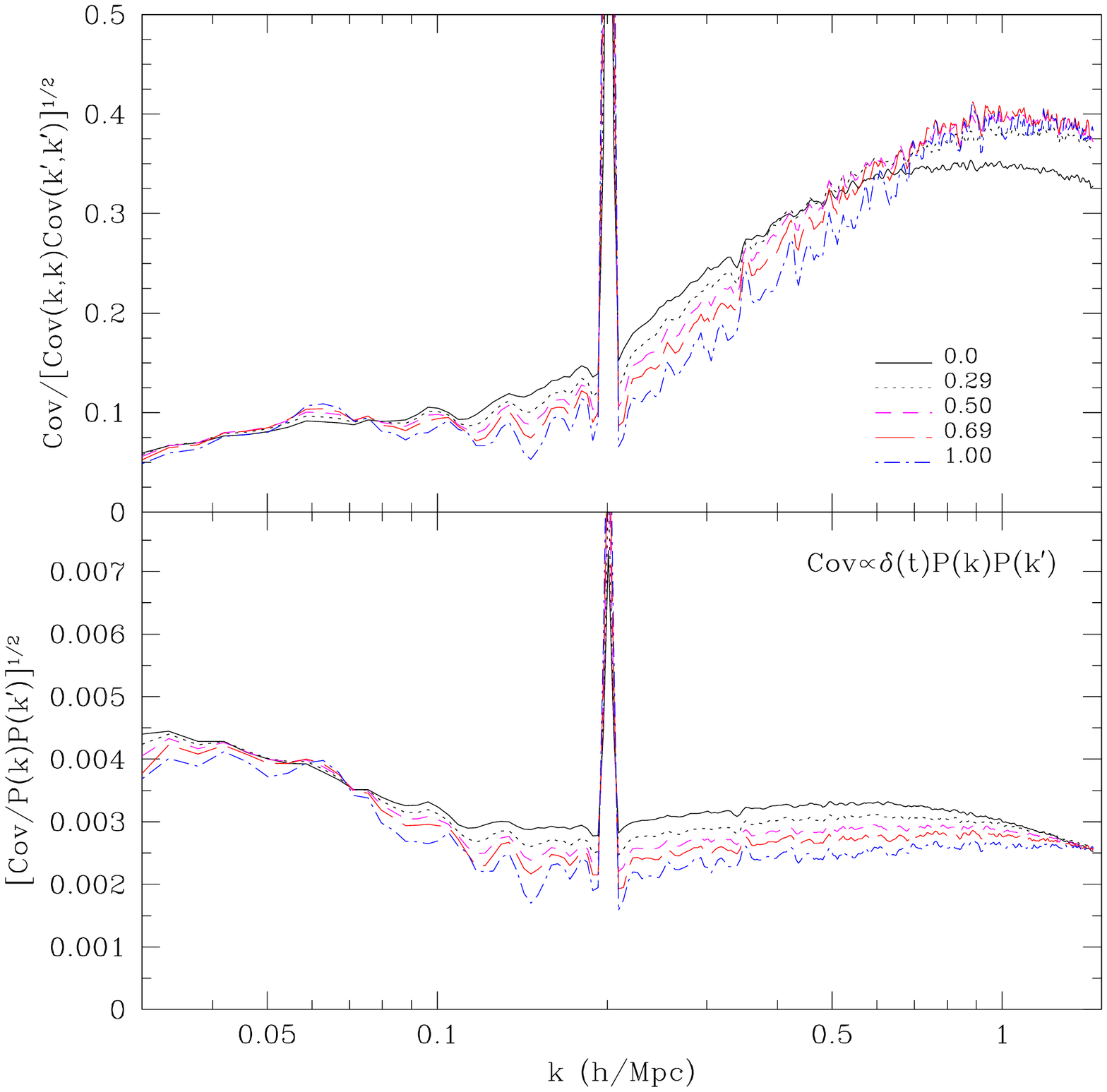}
\caption{Evolution of the covariance matrix with redshift for
  $k^\prime=0.2\,{\rm h\,Mpc}^{-1}$ in the A1.5 simulations. Top and bottom panels
  show the covariance coefficient and the covariance matrix normalised to the power spectra. 
  Curves on the
  right panels were scaled up with the linear growth factor
  $\delta(t)$. Results indicate that at very long waves
  $k\lsim 0.07\,{\rm h\,Mpc}^{-1}$ the covariance matrix grows very
  fast as $C\propto \delta(t)P(k,t)P(k^\prime,t) \propto
  \delta^5(t)$. At intermediate wave-numbers
  $0.1\,{\rm h\,Mpc}^{-1} < k < 0.5\,{\rm h\,Mpc}^{-1}$ the growth is even faster
   $C\propto \delta^{5.25}(t)$.}
\label{fig:CovZB}
\end{figure*}

\makeatletter{}\begin{figure*}
\centering
\includegraphics[width=0.49\textwidth]
{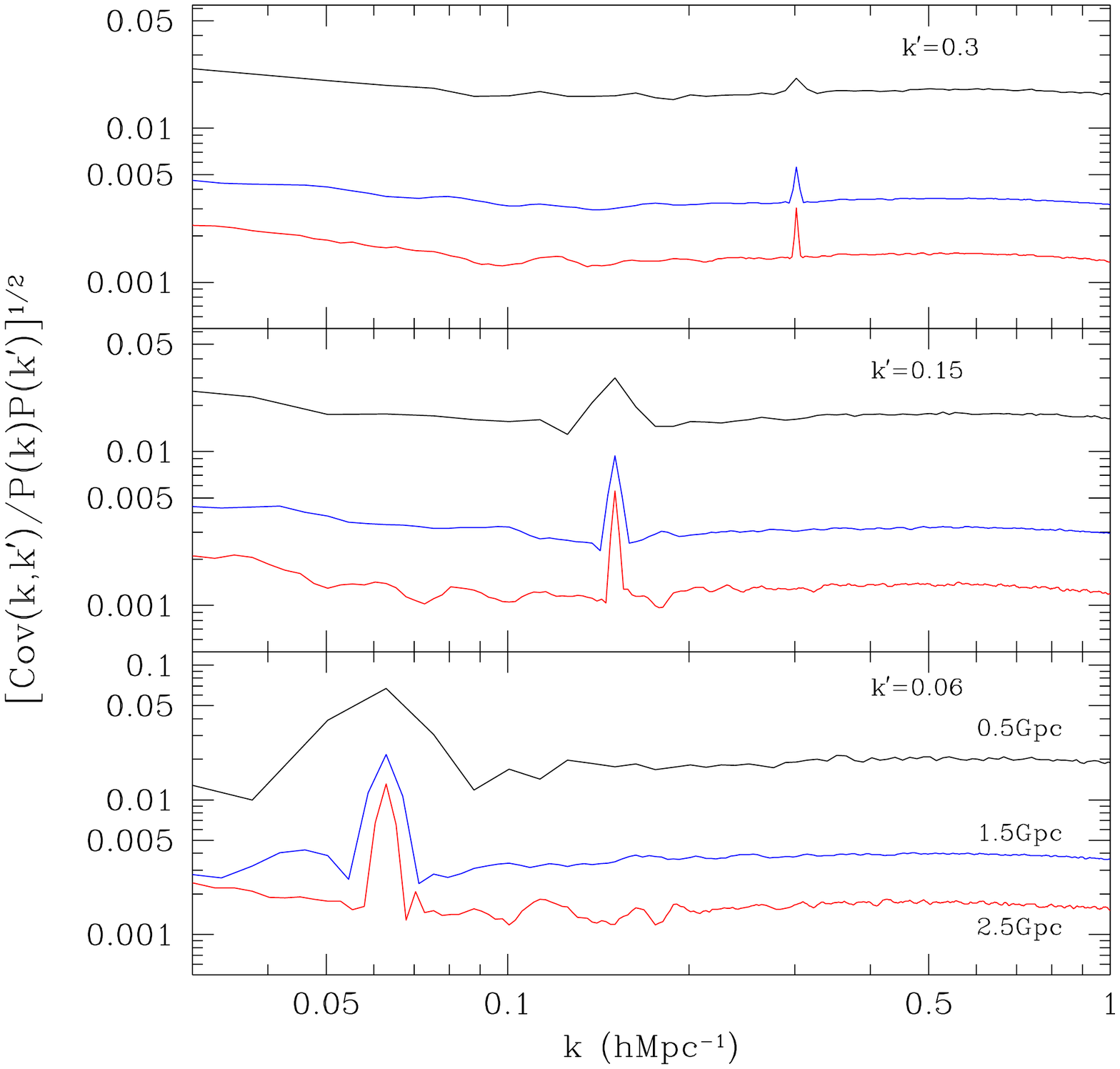}
\includegraphics[width=0.49\textwidth]
{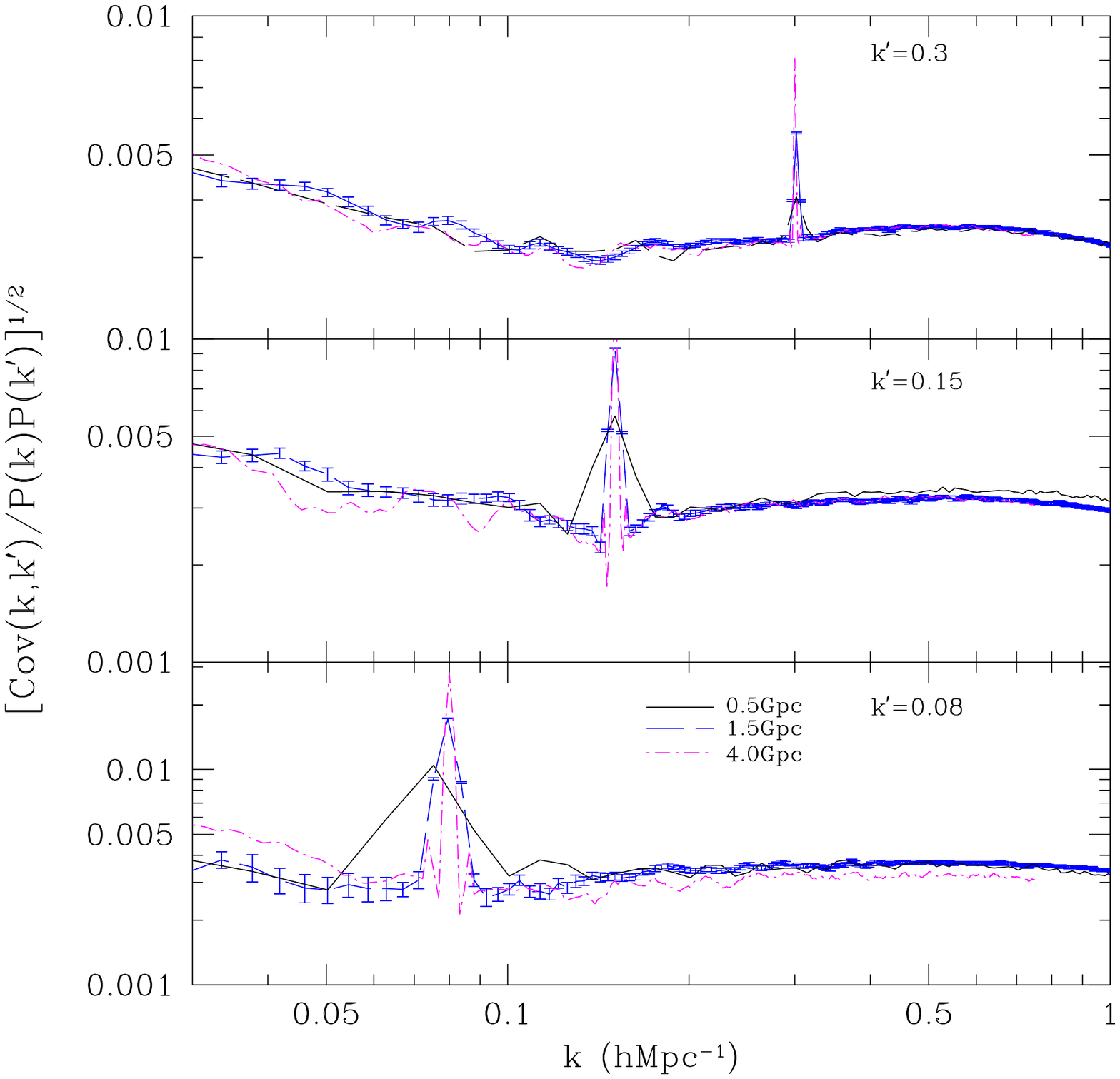}
\caption{Dependance of the covariance matrix on box size. PPM-GLAM simulations
  with three box sizes are used: $L=0.5\Gpch$\, (A0.5), 1.5\Gpch\,
  (A1.5), and 2.5\Gpch\, (A2.5). In all cases the binning of the power
  spectra is defined by the fundamental harmonic $\Delta k =2\pi/L$
  resulting in the finest binning and largest noise in the biggest
  box. {\it Left:} The unscaled covariance matrix dramatically declines
  with increasing $L$.  {\it Right:} Covariance matrices of the A0.5, A1.5
  and A4.0 simulations are scaled to the same volume of $1.5\Gpch$.
}
\label{fig:CovZ0box}
\end{figure*}

Figures~\ref{fig:CovZ0} and \ref{fig:CovZB} demonstrate that the
non-diagonal terms of the covariance matrix have a complex
structure. They depend on both $k$ and $k^\prime$ in a non-trivial way
and they evolve with redshift in a complicated fashion. The bottom
left panel in Figure~\ref{fig:CovZ0} highlights one of the main issues
with the non-diagonal terms: each term is small $\sim 3\times 10^{-3}$
as compared with the signal, but there are many of them. The fact
that individual components are so small is one of the reasons why one
needs thousands of realisations to estimate the covariance matrix
reliably.

However, with the exception of the diagonal componentes, the
covariance matrix is a smooth function that can be approximated
analytically using a relatively small number of parameters.  The
following approximation gives 3\% accuracy for the $z=0$ covariance matrix
at $k\gsim 0.1\,h{\rm Mpc}^{-1}$ and $\sim 10\%$ at smaller wave-numbers:
\begin{eqnarray}
C(k,k^\prime)&=&8.64\times 10^{-6}L^{-3}P(k)P(k^\prime)  \nonumber\\
                 &\times&\left[1 + g(k)+ g(k^\prime)  \right]^2 \\
                 &\times&\left[1 -\alpha e^{-\frac{(k-k^\prime)^2}{2\sigma^2}} 
                    +(1+\alpha) e^{-\frac{(k-k^\prime)^2}{2(0.1\sigma)^2}}\right]^2, \nonumber \\ \nonumber
\label{eq:fit1}
\end{eqnarray}
where
\begin{equation}
         g(k) = 1.4\exp\left[-\left(\frac{k}{0.07}\right)^2\right] 
               +\frac{1.1k^{0.60}}{1+1.2k^3},  
\label{eq:fit2}
\end{equation}
and 
\begin{equation}
        \alpha=0.03/k^\prime,\qquad \sigma= 4\kappa =4(2\times 10^{-3}\pi/L).
\label{eq:fit3}
\end{equation}
Here the box size $L$ is in units of $\Gpch$ and wave-numbers are in
units of $h\,{\rm Mpc}^{-1}$. In spite of the fact that the fit has 10
free-parameters, it is still a useful approximation.

The approximation for $C(k,k^\prime)$ is so complicated because the
covariance matrix has a complex dependence on $k$ and $k^\prime$. For
a fixed $k^\prime$ the covariance matrix declines with increasing
$k$. For example, the covariance matrix declines by a factor of
$\sim 2.5$ from $k= 0.03\,h{\rm Mpc}^{-1}$ as compared to
$k= 0.1\,h{\rm Mpc}^{-1}$. It reaches a flat minimum at
$k\approx (0.1-0.2)h\,{\rm Mpc}^{-1}$ and then increases by factor
$~1.5$ and reaches a maximum at $k\approx (0.5-0.6)h\,{\rm Mpc}^{-1}$.
In addition, it has a dip on both sides of the diagonal components
(see Figure~\ref{fig:CovZ0}), which is approximated by the last terms in eq.~(\ref{eq:fit1}).

The approximation given by eqs.~(\ref{eq:fit1}-\ref{eq:fit3}) are only for the
non-diagonal components of $C(k,k^\prime)$. For the diagonal
components the fallowing approximation gives a $\sim 2\%$-accurate fit
to the A0.9 simulations:
\begin{equation}
C(k,k) = P^2(k)\left[ \alpha\frac{2}{N_h} +A^2\right],
\label{eq:DiagFit}
\end{equation}
where parameter $A$ is a slowly increasing function of wave-number:
\begin{equation}
A = \frac{5.5\times 10^{-3}}{L^{3/2}}\left[1 +\left(\frac{k}{1.5}\right)^{1/2}-\frac{k}{4}\right], 
   \quad k<1\,h{\rm Mpc}^{-1}.
\label{eq:DiagFitA}
\end{equation}

The evolution of the covariance matrix with redshift is equally
complex as that illustrated in Figure~\ref{fig:CovZB} that shows
$C(k,k^\prime)$ for the A1.5 simulations at different
redshifts. Curves on the right panels were scaled up with the linear
growth factor $\delta(t)$. Results indicate that on very long waves
$k\lsim 0.07\,{\rm h\,Mpc}^{-1}$ the covariance matrix grows very fast
as $C\propto \delta(t)P(k,t)P(k^\prime,t) \propto \delta^5(t)$.  At
the intermediate scales $0.1\,{\rm h\,Mpc}^{-1} < k < 0.5\,{\rm h\,Mpc}^{-1}$
the growth is even faster: $C\propto \delta^{5.25}(t)$

  We also conclude that the covariance matrix decreases with the
  increasing computational volume. We already saw it for the diagonal
  components. The same is true for the non-diagonal terms as
  illustrated by the left panels in Figure~\ref{fig:CovZ0box}. The
  scaling of the whole covariance matrix with volume is trivial:
  $C\propto L^{-3}$. It is not an approximation or a fit. It is a
  scaling law. Right panels in the figure show the results for $C$
  rescaled to the $(1.5\Gpch)^3$ volume of the A1.5 simulations. This
  is an important scaling, which is often forgotten. Errors in the
  estimates of the power spectrum of fluctuations (and thus the errors
  in cosmological parameters) would have been too large if one were to
  stack together many small simulations to mimic large observational
  volume {\it and forget to re-scale the covariance matrix}. However,
  when the rescaling is done, the small-box simulations perfectly
  reproduce the clustering signal and statistics of much larger
  simulations.

\makeatletter{}\section{Comparison with other results}

%\subsection{Code Efficiency}
The speed of \Nbody\ simulations is very critical for the generation of mock
galaxy samples -- the ultimate goal for the GLAM project. Computational timings of
some PM codes are available \citep{Izard2015,Feng2016,Koda2016}, but
those timings are performed for different configurations, i.e. different
number of particles, mesh-sizes, time-steps, and computing facilities.
Fortunately, rescaling of the timings can be done because most of the
results are relatively recent (and use similar Intel processors). Most of
the CPU is used for numerous FFTs. So, results can be rescaled to the
same configuration. We re-scale all relevant numbers to a hypothetical
simulation with $\Np =1000$, $\Ng=3000$, and the number of time-steps
$\Ns=150$. Table~\ref{table:compare} presents the CPU and memory
required to run such a simulation with different codes.

  \makeatletter{}\begin{table}
    \caption{Rescaled CPU and memory required by different PM codes to 
make one 150 time-steps simulation with 1~billion particles and $\Ng=3000$ mesh. }
\begin{tabular}{ l | c | c | c |l }
\hline  
 & GLAM &  COLA$^1$  & ICE-COLA$^2$   & FastPM$^{3}$  
\tabularnewline
CPU, hrs & 146 & 220 & 320 & 567
\tabularnewline
Memory, Gb & 123 & 240 & 325 & 280
\tabularnewline
\hline
\multicolumn{5}{l}{\quad  $^1$\citet{Koda2016}, $^2$\citet{Izard2015}, 
     $^3$\citet{Feng2016}}
\tabularnewline
\end{tabular}
\label{table:compare}
\end{table}

So far PPM-GLAM code is the fastest available code for this type of
simulations: it is a factor of 2-3 faster than some other codes and it
requires significantly less memory. This is not surprising considering
GLAM's simplicity and efficiency.  The code is particularly tuned for
production of thousands of realisations. Indeed, it is trivially
parallelised with MPI library calls to run hundreds of realisations at
a time, which is exactly what we have done. It also has an advantage
that it can be used on a small cluster or on a single computing node
without installed MPI library. It has a disadvantage that it is
limited by the memory available on a single node.

It is interesting to note that approximate methods for the generation of
a large number of galaxy/halo mock catalogs \citep[see][]{Chuang2015} , while 
promising fast and cheap production of thousands
of catalogs, may not be very efficient and fast as one may
imagine. For example, \citet{Kitaura2016} generated $12,288$ mock
catalogs for the BOSS galaxy survey \cite{Kitaura2016} using the PATCHY-code
\citep{Kitaura2013}, which is based on a combination of
second-order perturbation theory and spherical infall model (ALPT). The
results are very valuable for the project and score really well when
compared with other methods as those discussed in \citet{Chuang2015}. The code 
uses a $960^3$ mesh, $2.5\Gpch$ box and spatial resolution of $2.6\Mpch$. It
produced $40,960$ realisations with 12.2~CPU-hours and 24~Gb RAM per realisation. That
seems to be very fast when one compares it with ICE-COLA which
requires 2.6~Tb of RAM and 1000~CPU hours
\citep{Izard2015}.

However, this is a misleading comparison because the resolution of the
ICE-COLA simulations is $0.25\Mpch$ -- ten times better than PATCHY,
and COLA does not use inaccurate spherical collapse model. In order to
make a fair comparison, we estimate the CPU time needed for PPM-GLAM
to make a realisation with the the same box size, $960^3$ mesh and
force resolution of $2.6\Mpch$ as in PATCHY. We assume that 40
time-steps would be sufficient for this resolution (see
eq.~(\ref{eq:Stability})) and use $(960/2)^3$ particles. The CPU time
to make one $N-$body simulation with PPM-GLAM is 1.4~CPU-hours, which
is smaller that for PATCHY. One simulation with
PPM-GLAM will also require about 4 times less memory than what is
needed for a PATCHY mock. In other words, it is faster and more
accurate to make one $N-$body simulation than use the approximate
PATCHY gravity solver.

Our results on convergence of the power spectrum are very similar to
those presented in \citet{Izard2015} for their ICE-COLA
simulations. Their Figure~6 indicates that 1\% error in the power
spectrum $P(k)$ is reached at $k\approx 0.7-0.8\,h{\rm Mpc}^{-1}$ for
a simulation with $\Delta x =0.25\Mpch$. Our
Figure~\ref{fig:ConvergePk} shows that 1\% error at
$k\approx 0.7\,h{\rm Mpc}^{-1}$ occurs for our PPM-GLAM simulation
with a similar resolution of $\Delta x =0.35\Mpch$. Convergence of the
power spectrum is somewhat worse for the FastPM code \citep{Feng2016}:
For the force resolution of $\Delta x =0.225\Mpch$ they reach 1\%
error in $P(k)$ only at $k\approx 0.3\,h{\rm Mpc}^{-1}$.  At
$k\approx 0.7\,h{\rm Mpc}^{-1}$ the error is 2\% for the FastPM
simulations. Note that it should have been other way around because
FastPM has a bit better force resolution and should have better
accuracy. It is not clear what causes this discrepancy.

Results on the covariance matrix of the dark matter power spectrum are
more difficult to compare because the covariance matrix depends
sensitively on cosmological parameters and has a complicated
shape. Some publications claim that, once the Gaussian diagonal terms
are removed, the covariance function is a constant:
$C(k,k^\prime)=\delta(k-k^\prime) \, C_{\rm Gauss}(k) +\sigma^2
P(k)P(k^\prime)$
\citep{Neyrinck2011,Mohammed2014,Carron2015}. This simple model is not
consistent with our results as presented in Figure~\ref{fig:CovZ0box}.
The main difference is the upturn of $C(k,k^\prime)$ at
$k< 0.1\,h{\rm Mpc}^{-1}$ where the covariance matrix changes quite
substantially. For example, $C(k,k^\prime)/P(k)P(k^\prime)$ changes by
a factor of two from $k^\prime=0.15\,h{\rm Mpc}^{-1}$ down to
$k^\prime=0.03\,h{\rm Mpc}^{-1}$. Even at larger wave-numbers there is
clear increase with increasing $k$, though this effect is not that large
($\sim 20\%$).

Recently, \citet{Blot2015} and \citet{Li2014} presented covariance matrices for
a large set of $N$-body simulations. It is difficult to compare our
results with those of \citet{Blot2015} because of the large differences in
cosmological parameters. However, the differences are smaller for
\citet{Li2014}. By comparing our covariance matrix for PPM-GLAM A0.5
simulations with \citet{Li2014}, as presented in Figure~1 of
\citet{Bertolini2016} and Figure~8 of \citet{Mohammed2016}, we find
that our results are within $\sim 10\%$ of \citet{Li2014}. However,
there are some differences when we use our full set of
simulations. For example, we clearly find the gradual increase in
$C(k,k^\prime)$ with increasing $k$ for $k> 0.1\,h{\rm Mpc}^{-1}$ and
subsequent decline (see right panel in
Figure~\ref{fig:CovZ0box}). The results of \citet{Li2014} are inconclusive
in this matter. 

\citet{Li2014} presented simulations only for a relatively small
$500\Mpch$ computation box. Instead, we study the covariance
matrix for vastly different box sizes, which allows us to asses effects of
SSC modes. We find very little effect of SSC on $C(k,k^\prime)$. 

\makeatletter{}\section{Conclusions}

Making accurate theoretical predictions for the clustering statistics
of large-scale galaxy surveys is a very relevant but complicated
physical process. In this paper we present and discuss our results on
a number of important aspects. Some of them are technical (e.g.,
convergence of the power spectrum) and others are of vital importance
for the properties of the dark matter density field that affects
directly the inference of cosmological parameters from large galaxy
catalogs (e.g. covariance matrices). There are different paths to
producing mock galaxy catalogs where the predictions for the dark
matter clustering and dynamics is a crucial stage of this
process. Only after the dark matter distribution is approximated one
way or another, and its clustering properties are reliable and
accurate, we can then build the next step of connecting dark matter
with galaxies, which will be subject of our forthcoming work.

The properties of the dark matter covariance matrix have been studied
using numerical simulations \citep{Takahashi2009,Li2014,Blot2015} or
analytical methods
\citep{Mohammed2014,Bertolini2016,Carron2015,Mohammed2016}.  Here we
present a detailed analysis of the covariance matrix based on a very
large set of \Nbody\ simulations that cover a wide range of numerical
and cosmological parameters. We study the structure, evolution, and
dependance on numerous numerical effects of the dark matter covariance
matrix using a new Parallel Particle-Mesh code, named
PPM-GLAM. Contrary to some previous results
\citep[e.g.,][]{Neyrinck2011,Mohammed2014} we find that the covariance
matrix $C(k,k^\prime)$ is a very complicated entity with complex
dependance on wave-numbers $k$ and $k^\prime$. We provide accurate
approximations in eqs.~(\ref{eq:fit1}-\ref{eq:DiagFitA}) of the
covariance matrix at $z=0$ for the standard $\LCDM$ model with Planck
parameters.

The covariance matrix evolves with redshift. It grows linearly for
very long waves: $C(k,k^\prime)/P(k)P(k^\prime) \propto \delta(z)$,
where $\delta(z)$ is the linear growth factor. At larger wave-numbers
it growths faster with
$C(k,k^\prime)/P(k)P(k^\prime) \propto \delta^{5/4}(z)$. The fast
growth of the covariance matrix implies that $C(k,k^\prime))$ must
depend on the overall normalisation $\sigma_8$ of the power spectrum
and very likely on other cosmological parameters. This is hardly
surprising considering that the power spectrum -- second order
clustering statistics -- depends on cosmology. Why the third-order
statistics such as the covariance matrix would not? 

We use vastly different simulation volumes to study the effects of SSC
waves -- waves longer than the simulation box. We clearly see these
effects in the power spectrum $P(k)$, but they occur only on very
small wave-numbers $k\lsim 0.03\,h{\rm Mpc}^{-1}$ and only for small
simulation boxes $L\lsim 500\Mpch$. There are no detectable SSC
effects for simulation boxes $L=1.5\Gpch$.

\makeatletter{}\begin{figure*}
\centering
\includegraphics[width=0.8\textwidth]
{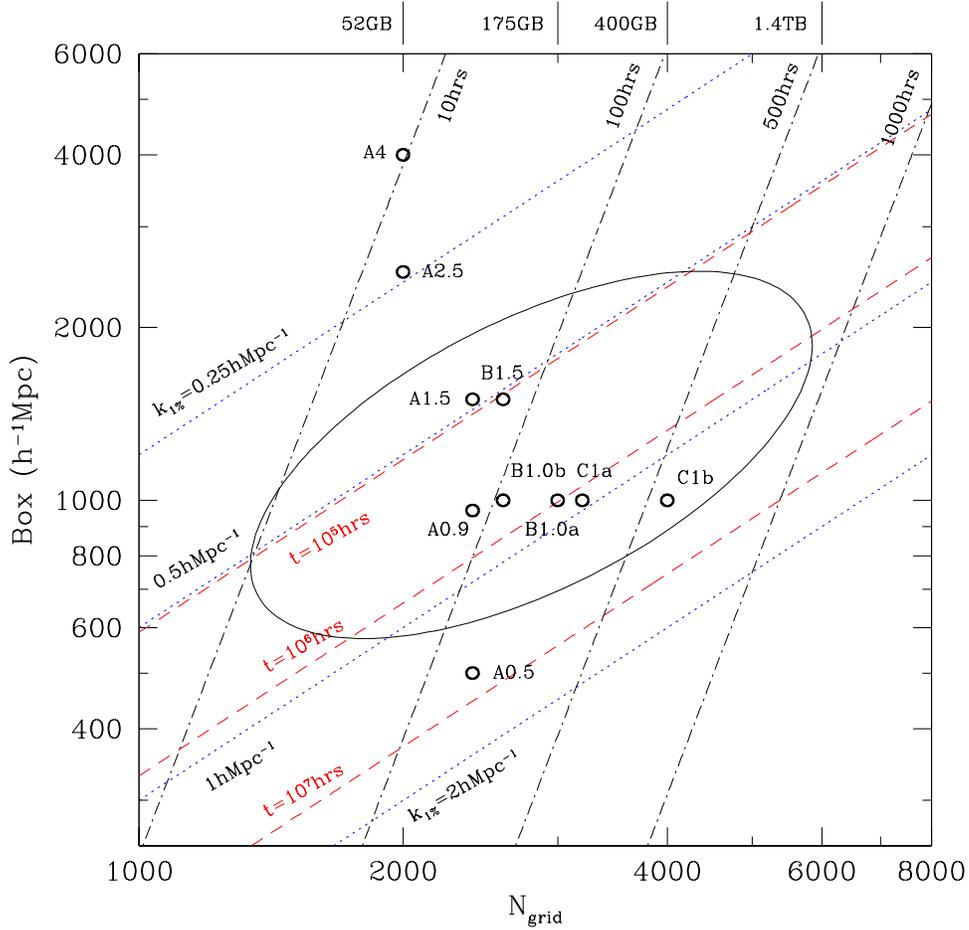}
\caption{Dependance of the different numerical  parameters of the PPM-GLAM
  simulations on box- and mesh-size. The vertical lines at the top-axis of the
  plot show the computer memory required for a simulation with mesh-size 
  $N_{\rm grid}$ and number of particles $\Np=N_{\rm grid}/2$.
  Dot-dashed lines correspond to the number of CPU-hours needed to make a
  single realisation with the given combination of mesh-size (and particles) and number of steps
  as defined by eq.~(\ref{eq:Nstot}). Diagonal dotted (blue) lines
  show constant values of $k_{1\%}$ (the wave-number at which the
  error in $P(k)$ reaches the level of 1\%). In order
  to achieve better resolution than a selected value of $k_{1\%}$, the simulation
  parameters (box- and mesh-size) should be set to those values located below the corresponding $k_{1\%}$
  dotted line. Dashed (red) lines are lines of constant CPU time (in hours) required to make
  a set of PPM-GLAM simulations with the cumulative volume of $5000\,(\Gpch)^3$.  In
  order to avoid large Super Sample Covariance (SSC) defects, the
  simulations should have large enough box-size $L>\gsim 500\,\Mpch$. The requirements to have
  a large number of realisations $N_{\rm r}$, for a given
  CPU time and accuracy, tend to reduce the simulation box-size. Overall, these different
  constraints tend to limit the selection of computational parameters
  to the oval area indicated in the plot. Open circles corresponds to the PPM-GLAM simulations
  listed in Table~\ref{table:simtable} and used in our analysis.}
\label{fig:Performance}
\end{figure*}

The optimum selection of the computational volume of \Nbody\
simulations to estimate covariance matrices requires special attention
because it potentially can change the accuracy of the covariance
matrix and power spectrum. It definitely affects the required CPU time
and requested memory. With increasing volume of observational samples,
one naively expects that the computational volume of an individual
mock catalog also must increase, and thus, cover the whole survey
volume \citep[e.g.][and references therein]{Monaco2016}. Yet, this is
not necessarily correct.  Weak-lensing samples still cover relatively
small volumes and in this case using simulations that cover the
observed samples may be vulnerable to large SSC effects. So, much
larger computational boxes may be required.

At the same time, the very large survey volumes of e.g. DESI
\citep{DESI2016} and Euclid \citep{Euclid} may not require equally
large simulation volumes.  We argue that one can replicate (stack) a
smaller but still significant simulation box as many times as needed
to cover the larger observational volume and on this way to produce a
viable mock galaxy catalog. As long as we are only interested in a
quantity (e.g., power spectrum or correlation function) that is
defined on scales significantly smaller than the computational volume,
there should be no defects. Indeed, Figure~\ref{fig:ConvergePk} shows
that the power spectrum is not affected by the simulation box. The
covariance matrix is not affected either after it is scaled down by
the ratio of the observational volume to the volume of the
simulation. Indeed, we already saw that the covariance matrix scales
proportionally to the volume (see, eqs.~(\ref{eq:fit1} -
\ref{eq:DiagFitA}) and Figure~\ref{fig:CovZ0box}).  For example, for
the analysis of BAO peaks and scales of $\sim 100\,\Mpch$ a simulation
box of size $L\approx (1-1.5)\Mpch$ is a good choice.

Finally, we estimate the computational resources -- CPU time and
computer memory -- required to run a large set of PPM-GLAM simulations for
different combinations of box size $L$, accuracy of the power
spectrum and mesh size $\Ng$. There are many factors that define
the optimal selection of computational parameters, including the number of
time-steps, the number of realisations, the effects of super-sample waves and
the limitations on the available computer memory.
Using the results presented in this paper on the  errors in the power
spectrum (see Figures~\ref{fig:ConvergePk} and \ref{fig:ConvergePMA})
we provide estimates for (i) the wave-number $k_{1\%}$ at witch the error in $P(k)$
reaches the level of 1\%:
\begin{equation}
  k_{1\%}= \frac{0.3}{\Delta x} = \frac{0.3\Ng}{L},
\end{equation}
where the box size $L$ is given in units of $\Mpch$. Lines of constant
$k_{1\%}$ are shown as dotted (blue) lines in
Figure~\ref{fig:Performance}. The larger the value of $k_{1\%}$,
better is the performance reached at smaller scales. (ii) to estimate
the number of time-steps $\Ns$\ required for a given simulation, we
assume that on average the particles should not move more than $1/3$
of a cell. This includes fast-moving particles in clusters of
galaxies. We assume that the $rms$ 3D-velocity for the particles in
galaxy clusters is $v\approx 2000\,\kms$.  Estimating the number of
steps as $a/\Delta a =\Ns$ and using eq.~(\ref{eq:Stability}) with
$\beta=1/3$, we find that the number of time-steps is
\begin{equation}
 \Ns\approx \frac{60\Ng}{L},
\label{eq:Nstot}
\end{equation}
 where the
box size is given in units of $\Mpch$.

Thus, the total amount of CPU-hours required for producing $N_r$
simulations with box size $L$ and mesh size $\Ng$ is

\begin{equation}
 t_{\rm tot} = N_r\Ns\Ng^3t_1 = 2.4\times 10^{-9}N_rN_g^4L^{-1},
\label{eq:ttotA}
\end{equation}
where $t_1$ is the CPU-hours required per time-step. Here we use
the timings provided in the first raw of Table~\ref{table:cpu}. It is also
convenient to use the total volume covered by all set of realisations, i.e.
$V=N_rL^3$.  Using the expressions for the total volume and CPU-hours, $V$ and $t_{\rm tot}$, we can
write a relation between the required box size $L$ and grid $\Ng$ to run many simulations covering a
volume $V$ under the condition that the total CPU is $t_{\rm tot}$:
\begin{equation}
L = \left( \frac{2.4\times 10^{-9}V}{t_{\rm tot}} \right)^{1/4}\Ng .
\label{eq:LNg}
\end{equation}

In Figure~\ref{fig:Performance} we plot lines of $L(\Ng)$ for a
somewhat arbitrary value of $V=5000(\Gpch)^3$ and for three different
CPU times of $10^5,10^6,10^7$CPU-hours. Additional constraints are
coming from the SSC modes that limit the box size, which we assume to
be bigger than $L=500\,\Mpch$. The number of realisations should be
large (thousands) regardless of the total covered volume.  This tends
to limit the box size to lower values, or, equivalently, to increase
proportionally the CPU time.  All theses limitations are shown in
Figure~\ref{fig:Performance}. They indicate that currently the
selection is limited to the parameters inside the oval drawn in the
plot.

The results presented in this work demonstrate that the Parallel
Particle-Mesh GLAM code, at a very low computational cost, generates
the accurate dark matter clustering statistics required for the
production of thousands of mock galaxy catalogs. The next step, which
will be presented in a future paper, will be to develop a bias scheme
that will take as input the accurate density field provided by
PPM-GLAM and produce those large galaxy catalogs in the context of the
upcoming and existing large-scale galaxy surveys.
  
\section*{Acknowledgements}

A.K. acknowledges support of the Fulbright Foundation and support of
the Instituto de Fisica Teorica, CSIC, Madrid,
Spain. F.P. acknowledges support from the Spanish MINECO grant
AYA2010-21231-C02-01. We thank Johan Comparat (IFT, Madrid), Claudia
Scoccola (IAC, Tenerife), and Sergio Rodriguez-Torres (UAM, Madrid)
for comments and fruitful discussions.  The PPM-GLAM simulations have
been performed on the FinisTarrae II supercomputer at CESGA in
Galicia, supported by the Xunta de Galicia, CISIC, MINECO and EU-ERDF.

\bibliography{PMP}
\bibliographystyle{mn2e}

\appendix
\section{The PPM-GLAM code}
\label{sec:AppendixA}

In general, a cosmological PM code consists of three steps to evolve
the particles: (1) Using the particle positions $\vc{r}$ to obtain the
density $\rho_{i,j,k}$ at the nodes of an homogenous 3D-mesh that
covers the computational domain, (2) Solve the Poisson equation on the
mesh, and (3) advance the particles to a new moment of time.

\subsection{ Density field:} We start with the calculation of the density field
produced by $N_{\rm p}^3$ particles on the $N_{\rm g}^3$ nodes of the mesh. In order
to assign the particle density to the 3D-mesh, we introduce a particle
shape \citep{HockneyEastwood}. If $S(x)$ is the density at distance
$x$ from the particle and $\Delta x$ is the cell size, then the
density at distance $(x,y,z)$ is the product $S(x)S(y)S(z)$. Two choices
for $S$ are adopted: Cloud-In-Cell (CIC) and Triangular Shaped Cloud
(TSC). Here we will use the CIC scheme, i.e.
\begin{equation}
  CIC:  S(x) = \frac{1}{\Delta x}\begin{cases} 1, \, |x|<\Delta x/2 \\ 0,
             \, {\rm otherwise} \end{cases}.
\label{eq:Scic}
\end{equation}
The fraction of particle mass assigned to a cell is just a product
of three weight functions $w(x)w(y)w(z)$, where
$\vc{r}=\vc{r_p}-\vc{x_i}$ is the distance between particles with
coordinates $\vc{x_p}$ and cell center $\vc{x_i}$. The weight
function is
$w(x)=\int^{x_i+\Delta/2}_{x_i-\Delta/2} S(x_p-x^\prime)dx^\prime$:

\begin{equation}
  CIC:  w(x) = \begin{cases} 1-|x|/\Delta x, \, |x|<\Delta x\\ 
             0, \phantom{mmmmm} {\rm otherwise} \end{cases}.\phantom{mmmmmmmm}
\label{eq:Wcic}
\end{equation}

Although these relations given in eqs.(\ref{eq:Scic}--\ref{eq:Wcic}) look
somewhat complicated, in reality they require very few operations in the
code. For the CIC scheme a particle contributes to the 8 nearest
cells. If the coordinates are scaled to be from $0$ to $N_{\rm g}$,
where $N_{\rm g}$ is the size of the grid in each direction, then
taking an integer part of each particle coordinate with center
$(x,y,z)$ - in Fortran: $i=INT(x) ...$ - gives the lower bottom grid
cell $(i,j,k)$.  Then, the distance of the particle from that cell
center is $dx =x-i, dy=y-j, dz=z-k$.

\subsection{Gravitational potential}
Having the density field $\rho_{i,j,k}$, we can estimate the
gravitational potential by solving the Poisson equation, which for clarity we simply write as
\begin{equation}
  \nabla^2\phi = 4\pi G\rho(\vc{x}).
\label{eq:Poisson}
\end{equation}

We start with applying a 3D Fast Fourier Transformation (FFT) to the
density field. That gives us the Fourier components on the same grid
$\tilde\rho_{\vc{k}}$, where $\vc{k}$ is a vector with integer
components in the range $0,1,\dots,N_{\rm g}-1$.  Now we multiply the
harmonics $\tilde\rho_{i,j,k}$ by the Green functions $G(\vc{k})$ to
obtain the Fourier harmonic amplitudes of the gravitational potential
$\phi$, i.e.

\begin{equation}
\tilde\phi_{i,j,k} = 4\pi G \tilde\rho_{i,j,k} G(\vc{k}),
\label{eq:GreenA}
\end{equation}
and then do the inverse FFT to find out the gravitational potential
$\phi_{i,j,k}$. Note that all these operations can be organized in
such a way that only one 3D-mesh is used -- no additional RAM memory is
required.

The simplest, but not the best, method to derive the Green functions
is to consider $\phi_{i,j,k}$ and $\rho_{i,j,k}$ as amplitudes of the
Fourier decomposition of the gravitational potential in the
computational volume, and then to differentiate the Fourier harmonics
analytically. This gives
\begin{equation}
         G_0(\vc{k}) = -\frac{1}{k_x^2+k_y^2+k_z^2} = 
              -\left( \frac{L}{2\pi}\right)^2\frac{1}{i^2+j^2+k^2}.
\label{eq:GreenK} 
\end{equation}
A better way of solving the Poisson equation that we use in our
PPM-GLAM code is to start with the finite-difference approximation of
the Laplacian $\nabla^2$. Here we use a the second order Taylor
expansion for the spacial derivatives:
\begin{eqnarray} 
 \nabla^2\phi &=& \frac{\partial^2\phi}{\partial x^2} +
\frac{\partial^2\phi}{\partial y^2} +\frac{\partial^2\phi}
        {\partial z^2}\nonumber \\ &\approx&
        [\phi_{i+1,j,k}-2\phi_{i,j,k}+\phi_{i-1,j,k} \\
    &+& \phi_{i,j+1,k}-2\phi_{i,j,k}+\phi_{i,j-1,k} \nonumber \\
    &+& \phi_{,j,k+1}-2\phi_{i,j,k}+\phi_{i,j,k-1}]/\Delta x^2. \nonumber
\end{eqnarray}
This approximation leads to a large system of linear algebraic equations:
$A\phi =4\pi G\rho$, where $\rho$ is the vector on the right hand
side, $\phi$ is the solution, and $A$ is the matrix of the
coefficients. All of its diagonal components are equal to -6, and all 6
nearest off-diagonal components are 1. The solution of this matrix
equation can be found by applying the Fourier Transformation. This
provides another approximation for the Green functions:
\begin{equation}
         G_1(\vc{k}) =  \frac{\Delta x^2}{2}\left[
                                        \cos\left(\frac{2\pi i}{N_{\rm g}}\right)
                                       +\cos\left(\frac{2\pi j}{N_{\rm g}}\right) 
                                       +\cos\left(\frac{2\pi k}{N_{\rm g}}\right)
                       -3\right]^{-1}.
\label{eq:GreenF} 
\end{equation}
For small $(i,j,k)$, eq.(\ref{eq:GreenF}) gives the same results as
eq.(\ref{eq:GreenA}). However, when $(i,j,k)$ is close to $N_{\rm g}$,
the finite-difference scheme $G_1$ provides less suppression for
high-frequency harmonics and thus gives a stronger and more accurate
force at distances closer to the grid spacing $\Delta x$.
\citet{HockneyEastwood} argue that this happens because the
finite-difference approximation partially compensates the dumping of short
waves that are related with the density assignment.

\subsection{Time-stepping} 
We write the particle equations of motions and the
Poisson equation, using the particle momenta $\vec{p}\equiv a^2\dot{\vec{x}}$, as follows
\begin{eqnarray}
    \frac{d\vc{x}}{da}  &=&  {\bf u}, \qquad {\bf u}\equiv \frac{{\bf p}}{a^3H}, 
                             \quad  \vec{p}\equiv a^2\dot{\vec{x}}, \label{eq:dx}\\
    \frac{d\vc{p}}{da}   &=&  {\bf g}, \qquad {\bf g}\equiv -\frac{\nabla\phi}{aH}, \label{eq:motion} \\
    \nabla^2\phi        &=&  \frac{3}{2}\frac{H_0^2\Omega_0\delta_{\rm dm}}{a}, \label{eq:Pois} \\
     H^2                &=& H_0^2\left(\frac{\Omega_0}{a^3}+\Omega_{\Lambda,0}\right), \quad   
                              \Omega_0 + \Omega_{\Lambda,0} =1.\label{eq:Friedmann}
\label{eq:Charac}
\end{eqnarray}
Here we specifically assumed a flat $\Lambda CDM$ cosmological model with the
cosmological constant characterised by the density parameter
$\Omega_{\Lambda,0}$ at redshift $z=0$. 

Because we start the simulations at a relatively high redshift
$z_i\approx 100$, and because the number of time-steps
$N_s\approx 100-200$ is not large, the time-stepping scheme should be
carefully selected and tuned.
In the sense of the accuracy of the time-stepping algorithms, there are
two regimes: (1) when fluctuations grow nearly
linearly at large redshifts and when the expansion factor $a$ may change substantially
over a single time-step, and (2) later moments when fluctuations are in
the non-linear regime with $a$ changing very little over a
time-step. Both regimes present a challenge for accurate simulations with a
small number of time-steps. There are different possibilities in handling
these challenges.  

At the linear stage of evolution the main concern and the main test is
the linear growth of fluctuations.  To address this problem COLA \citep{COLA,Tassev2015}
splits the changes in coordinates into a second-order perturbation term
(estimated by a separate algorithm), and a residual, which is
integrated using \Nbody\ methods. Instead, QPM \citep{QPM} uses a
logarithmic time-step (constant in $\Delta a/a$). COLA's
time-stepping is a good idea (but very expensive) for the quasi-linear
regime. However, at the very nonlinear stages of evolution, when the second
order perturbation approximation is bound to be not valid, the
splitting of the coordinate advances into two terms that cannot produce any
benefits, and thus, it seems to be just a waste of CPU. At this stage
a constant-step leap-frog scheme is preferred: it is time-symmetric,
second-order accurate and hamiltonian preserving approximation.

Motivated by these considerations, we select a time-stepping scheme
which uses a constant time-step at low redshifts $z< z_{\rm limit}$,
but periodically increases the time-step at large redshifts
$z> z_{\rm limit}$. The parameter $z_{\rm limit}$ defines the
transition from early quisi-linear to late non-linear regimes. With a
resolution of $\Delta x=(0.3-0.5)\Mpch$ in our simulations, some halos
may start to collapse and virialise at $z< z_{\rm limit}$. This
is the stage when we switch the time-stepping to the normal leap-frog scheme
with a constant time-step.  For our simulations we select $z_{\rm limit}=3$.

{\it (i) Early stages of evolution $z> z_{\rm limit}$.}  It is important
to estimate how the terms $\vec{u}$ and $\vec{g}$, in the
right-hand-sides of equations~(\ref{eq:dx}-\ref{eq:motion}), evolve
with the expansion parameter $a$ at the linear regime. Because there
are terms with large powers of $a$, one may be concerned with the
accuracy of the integration of quickly evolving terms. However, when one
considers all the terms, the situation is much less alarming. Indeed,
in the linear regime the peculiar gravitational potential $\phi$ does
not change with $a$, and along the particle trajectory
${\bf g}(a) \propto a^{1/2}$, leading to ${\bf p}\propto a^{3/2}$ and
${\bf u}\propto a^0$(constant). This means that there are no quickly
evolving terms in the equations of motions. This slow evolution of the
${\bf u}$ and ${\bf g}$ terms allows one to periodically increase the
time-step without substantial loss of accuracy. We do it by testing
the magnitude of $\Delta a/a$. If this ratio falls below a specified
value $(\Delta a/a)_{\rm min}$ (typically $3-5\times 10^{-2}$), the time-step
$\Delta a$ is increased by factor 3/2. 

We can write the time-stepping scheme using a sequence of kick $K$ and
drift $D$ operators, which are defined as
advances of particle momenta $\vc{p}$ and particle coordinates
$\vc{x}$ from moment $a$ to moment $a+\Delta a$:
\begin{eqnarray} 
K(\Delta a,a,\tilde a) &:& {\bf p}(a+\Delta a) = {\bf p}(a) + {\bf g}({\tilde a})\Delta a, \\
D(\Delta a,a,\tilde a) &:& {\bf x}(a+\Delta a) = {\bf x}(a) + {\bf u}({\tilde a})\Delta a, 
\label{eq:KD}
\end{eqnarray}
where $\tilde a$ is the moment at which either $\bf u$ or $\bf g$ are estimated.

If we start with particle momenta at the time $a_{-1/2}=a_0-\Delta a/2$ (a half time-step behind
the coordinates defined at $a_0$), and use the notation $a_m=a_0+m\Delta a$, the standard
leap-frog scheme can be written as the following sequence of kick and drift operators:

\begin{eqnarray} 
&&K(\Delta a,a_{-1/2},a_0)D(\Delta a,a_0,a_{1/2})\\
&&K(\Delta a,a_{1/2},a_1)D(\Delta a,a_1,a_{3/2})\\
&&K(\Delta a,a_{3/2},a_2)D(\Delta a,a_2,a_{5/2})\dots
\label{eq:LeapFrog}
\end{eqnarray}

When at some moment $a_0$ we need to increase the time-step by factor 3/2, we
do it by making a stronger kick and then by modifying the time-step to
the new value of $\Delta a^\prime = 3\Delta a/2$:
\begin{equation} 
K(5\Delta a/4,a_{-1/2},a_0)D(3\Delta a/2,a_0,a_{3/4})\dots
\label{eq:LeapFrog2}
\end{equation}
After applying the first pair of kick-drift operands, the normal setup
of the leap-frog scheme is restored with the particles momenta behind
the coordinates by a half of the new time-step. The code continues
the integration of the trajectories with a constant time-step until the moment
when $\Delta a/a$ becomes smaller than the minimum value. The
time-step is increased again by the factor 3/2, and the process
continues.

The truncation error for the variable step scheme can be found
similarly to the way how it is done for the standard leap-frog scheme by
eliminating the velocities from the scheme, and then by expanding
the coordinates in the Taylor series around moment $a_0$.  This gives
$x_{3/2}-(5/2)x_0+(3/2)x_{-1} = (15/8)g_0\Delta a^2$, and the
truncation error $\epsilon$ at the moment of time-step increase $a_0$ is:
\begin{equation} 
\epsilon = \frac{5}{16}{\dot g}_0\Delta a^3,
\label{eq:LeapFrog2Error}
\end{equation}
which should be compared with the truncation of the constant step leap-frog scheme:
\begin{equation} 
\epsilon = \frac{1}{12}{\ddot g}_0\Delta a^4.
\label{eq:LeapFrogError}
\end{equation}
The truncation error at the moment of modifying the time-step is
clearly larger than for the constant-step leapfrog, but it is still a
third-order approximation. The reason for that is the selection of the
numerical factor 5/4 in the kick operator (eq.~\ref{eq:LeapFrog2}), which kills the
second-order error. These errors are only for a single time-step. The
cumulative error for a large number of steps depends on how
single-steps errors accumulate. This typically results in scaling the force resolution
$\epsilon \propto \Delta a^2$ for the constant time-step. Because
there are only very few number of times when the time-step is
increased in our code (typically 5-10 times for the total $\sim 150$
of steps), the final error is mostly dominated by the cumulative error
of the constant-step kicks and drifts.

{\it (2) Late stages of evolution $z< z_{\rm limit}$.}  As fluctuations
evolve and become very nonlinear, halos start to form, merge and
grow. At this stage the main concern is how accurately the code traces
the evolution of dark matter in halos. The number of time-steps is an
important factor defining the accuracy. However, the number of steps
is just one of the factors: one cannot really find out the required number of
steps without specifying the force resolution and without knowing the science
application and requirements of the simulations.

Our goal is to generate PPM-GLAM simulations that reproduce the dark matter
density and velocity fields with the resolution of up to
$\sim 1/3-1/2$\,$\Mpch$. Peculiar velocities are an integral part of the
process implying that redshift distortions should be simulated, not
added posteriorly using some analytical prescription. The force
resolution and the magnitude of the peculiar velocities set stringent
constraints on the allowed time-step.

The largest peculiar velocities $\sim 1000-3000\,\kms$ occur in
clusters and large galaxy groups. The time-step should be small enough
so that per time-step a particle should move by less than a fraction
of a cell. Thus, for both stability and accuracy of the integration
\citep{HockneyEastwood},
\begin{equation}
\beta \equiv \frac{v\Delta t}{\Delta R} \lsim 1,
\label{eq:dtcondition}
\end{equation} 
were $v$ is the particle
velocity, $\Delta t$ and $\Delta R$ are the time-step and the (proper)
cell size. Assuming that the time-step is small, we can write
$\Delta t =\Delta a/aH(a)$. If $\Delta x =\Delta R/a$ is the comoving
cell size, then we can write $\beta$ in the following form:
\begin{equation}
  \beta = \frac{v}{a\Delta x}\frac{\Delta a}{aH(a)} = 
          \frac{v}{\Delta x H_0}\frac{\Delta a}{a}\sqrt{\frac{a}{\Omega_0+\Omega_\Lambda a^3}}.
\end{equation}
Scaling velocities and resolution to some characteristic values we finally
write the condition for selecting the time-step as follows
\begin{equation}
   \beta = 10 \left[\frac{\Delta a}{a}\right ] \left[\frac{v_{1000}}{\Delta x_{\rm Mpc}}\right ] 
               \sqrt{\frac{a}{\Omega_0+\Omega_\Lambda a^3}} < 1,
\label{eq:Stability}
\end{equation}
where $v_{1000}$ is the peculiar velocity in units of $1000\,\kms$ and
$\Delta x_{\rm Mpc}$ is the comoving cell size in units of $\Mpch$.

This condition is difficult to satisfy if the number of steps is
small. To make an estimate, let's assume that a PM code makes 40 time-steps using a constant-step
leapfrog scheme \citep[e.g. ICE-COLA and FastPM][]{Izard2015,Feng2016}. This implies that at
$z\approx 0$ the time-step is about
$\Delta a/a\approx 2.5\times 10^{-2}$.  Because we want the code to attain
realistic velocities inside clusters of galaxies, we take
$v =2000\,\kms$. For typical force resolution of $\Delta x=0.3\,\Mpch$ we find that
$\beta =1.7$. In other words, dark matter particles are moving too
fast for this combination of peculiar velocity and resolution. 

What happens if the time-step is too big? In this case large halos will not be as
dense as they should be and random velocities are smaller in the
central halo regions. This will be observed as a decline in the power
spectrum of dark matter. For example, \citet{Feng2016} using FastPM find a decline
of 4\% in the power spectrum at $k=1\Mpch$ for simulations with force
resolution $\Delta x=0.22\,\Mpch$ and 40 time-steps. However, the main
concern and the main problem is that the defect depends on the local density and $rms$
velocity. As such, it affects much more massive clusters, where velocities are
large, than small halos with small $rms$ velocities. 

In our simulations the
time-step at later moments becomes relatively small with a typical
value of $\Delta a/a\approx (0.75-1)\times 10^{-2}$, which is
sufficient even for fast particles in very massive clusters.

\subsection{Parallelization}

Parallelization of PPM-GLAM is done with OpenMP directives. Because
OpenMP can be applied only to memory on a single computational node,
this limits the number of particles and force resolution. This also
makes the code faster because the code does not use slow
communications across and iside computational nodes required for MPI
parallelization. Using only OpenMP directives also makes the code
simple and easy to modify. The later is very important because data
analysis routines are still being modified and improved.

For solving the Poisson equation the PPM-GLAM uses FFT fortran-90
routines for the real-to-real transformations provided by publicly
available code {\sc FFT5pack} \citep{FFT5}. This makes the code
portable: no libraries should be installed. Using MKL routines
provided by the Intel Fortran compilers may  further improve the
code performance.

Each OpenMP thread handles $\Ng^2$ 1-D FFT transformations. After
sweeping the 3-D mesh in two directions the matrix is transposed to
improve the data locality. Then the FFT is applied again twice: to
complete the 3-D sweep and to start the inverse FFT. The matrix is
transposed back, and the other two FFT sweeps are completed.  OpenMP
{\sc Atomic} directives are applied for density assignments, which
slows down the code, but allows it to run in parallel. The motion of
particles is a naively parallel part of the code. Overall, the code
uses only one 3-D matrix and requires three 3-D FFT passes.

\subsection{Performance of the PM code}
\makeatletter{}\begin{table*}
 \begin{minipage}{16.cm}
   \caption{Timing of the PPM-GLAM code for different computational systems.
     The columns give: (1) number of particles $N_p$, (2)
     number of grid cells $N_g$, (3) processor type and number of cores, (4)
     the total wall-clock time per step in minutes, (5) wall-clock
     time for the Poisson solver in minutes, (6) advancing particles timing in minutes and (7)
     density assignment timing in minutes. Columns (8--10) give the parameters $A, B, C$ for
     CPU time per cell and per particle in eq.~(\ref{eq:cpu}) in units of
     $10^{-8}$. Other columns provide: (11) CPU time per step in minutes, (12) CPU time per step per particle
     in $10^{-6}$ seconds, (13) CPU time in hours for a  un with 150 time-steps.}
\begin{tabular}{ l | c | c | c |  c|  c | c | c | c| c | c | c |l }
\hline  
 (1)   & (2)    &  (3)        & (4)       & (5)     & (6)       & (7)     & (8)   &  (9)  & (10)            & (11)      & (12)          & (13) 
\tabularnewline
 $N_p$ &  $N_g$ &  Processor  & Total     & Poisson & Particles & Density &  A    &   B   &  C              & CPU       & CPU           & CPU
\tabularnewline
       &       &  cores      & min   & min     & min       & min    &  & &         & step    & particle      & run
\tabularnewline
  \hline 
1200$^3$ & 2400$^3$ & Intel E5-2680v4 & 1.20 & 1.02 & 0.07      & 0.10   & 12.4  & 6.8  & 9.8             & 33.6     & 1.17         &  84
\tabularnewline
         &          & 2.4GHz 2x14     &
\tabularnewline
1200$^3$ & 2400$^3$ & Intel E5-2680v3 & 1.44 & 1.22 & 0.08      & 0.14   & 15.0  & 6.7  & 11.7              & 34.5     & 1.20         &  86
\tabularnewline
         &          & 2.5GHz 2x12     &
\tabularnewline
500$^3$ & 1000$^3$ & Intel E5-2680v4 & 0.075 & 0.062 & 0.0039 & 0.0088  & 10.4  & 5.2  & 11.8              &  2.1     & 1.01         &  5.2
\tabularnewline
         &          & 2.4GHz 2x14     &
\tabularnewline
1000$^3$ & 2000$^3$ & Intel E5-2680v4 & 0.65  & 0.55   & 0.037 & 0.057   & 11.5  & 6.2  & 9.6              &  18.2     & 1.09         &  45.5
\tabularnewline
         &          & 2.4GHz 2x14     &
\tabularnewline
1300$^3$ & 2600$^3$ & AMD 6174        & 2.23  & 1.72   & 0.21 & 0.30    & 28.2  & 27.5  & 39.3               &  107     & 2.92         &  267
\tabularnewline
         &          & 2.4GHz 4x12    &
\tabularnewline
1600$^3$ & 3200$^3$ & AMD 6376        & 2.44  & 2.11   & 0.11 & 0.22    & 24.7  & 10.3  &  20.6              &  156     & 2.28         &  390
\tabularnewline
         &          & 2.3GHz 4x16    &
\tabularnewline

\hline
\end{tabular}
\label{table:cpu}
\vspace{-5mm}
\end{minipage}
\end{table*}

The PPM-GLAM code was tested on a variety of processors (both Intel and
AMD) and computer platforms. Results of code timing are given in
Table~\ref{table:cpu} for different hardware configurations and
parameters of the simulations. As might have been expected, the Intel
processors are about twice faster than AMD when timing results are scaled to
CPU-hours per core. However, this is somewhat deceiving because the
AMD processors provide about twice more cores. If this is taken into
account, then the difference between AMD and Intel processors becomes
smaller.  For example, a single computational node with four AMD-6376
processors has the the same performance as a node with two Intel E5-2680v4
processors when rescaled to the same computational task.

Column~12 in Table~\ref{table:cpu} provides the CPU time scaled per individual particle. In that
respect it is a measure of the efficiency of the parallelisation and performance of our PM code. It shows 
that within $\sim 20$\% the code scales well for different number particles and mesh sizes. 

The computational cost of a PPM-GLAM simulation depends on the number of time-steps
$N_s$, the size of the 3D-mesh $N_{\rm g}^3$, and the adopted number of particles
$N_{\rm p}^3$. The CPU required to solve the Poisson equation is mostly
determined by the cost of performing a single 1D-FFT. Thus, it is proportional to
$(N_g\log{\Ng})^3$. There are some additional costs (e.g., two 3D-matrix transpositions 
to align the mesh with the memory of individual computational processors), but those are relatively small.

The required memory for a simulation is given by eq.~(\ref{eq:mem}). We
could have used double precision accuracy for the coordinates, as
adopted in FastPM by \citet{Feng2016}, but our estimates show that the
loss of coordinates accuracy at the edge of the simulation box are
practically negligible. For example, for an extreme configuration of a
$1000\,\Gpch$ simulation box with $\Ng=3000$ mesh particles moving for
13\,Gyrs with a constant drift velocity of $500\,\kms$, and with
additional random velocity of $1000\,\kms$, will have an error of just
$2\times 10^{-4}\Mpch$. This is very small uncertainty as compared
with the simulation cell size of $0.33\,\Mpch$.

While the CPU speed and RAM memory estimates are very favorable for a very
large number of medium-resolution PPM-GLAM simulations,
equations~(\ref{eq:cpu}-\ref{eq:mem}) clearly indicate that increasing
either the resolution or losing resolution for some code parameter
configurations can have serious repercussions. For example, increasing
the force resolution $\Delta x= L/\Ng$ by a factor of two, increases
the computational CPU cost eight times, i.e. a very large factor. Thus, the parameters of
the simulations should be selected very carefully. Loss of resolution may
happen as a side-effect of a modification in algorithms that at first
sight seems reasonable.

For example, the QPM code \citep{QPM} uses the Green functions given
in eq.~(\ref{eq:GreenK}) instead of the more advanced
eq.~(\ref{eq:GreenF}) \citep{HockneyEastwood} adopted in PPM-GLAM. Our
tests show that this change alone reduces the force resolution by
about 20 percent, which seems like a small loss, but not for a
cosmological PM code. In order to recover the loss, one would need to
increase the CPU and memory by a factor $1.7$. Because PM codes tend
to run at the limit of available computer memory, this factor
represents a serious disadvantage. One may also think of improving the
PM code, for example, by increasing the order of the gravitational
potential interpolation scheme \citep[QPM;][]{QPM} or by replacing
numerical differentiation by obtaining acceleration in the
Fourier-space \citep[COLA;][]{COLA}. Yet, higher-order schemes will
effectively reduce the resolution, and when compared at the same
resolution, these modifications only slow down the code without
gaining numerical accuracy.

One may  try to avoid numerical differentiation of the
gravitational potential by solving the acceleration in Fourier-space,
as done in the COLA and FastPM codes
\citep{COLA,Feng2016}. Potentially, that strategy could increase the
resolution, but it is not clear whether this procedure actually is
beneficial\footnote{We compare the errors in the power spectra at
  $k=0.3h{\rm Mpc}^{-1}$ shown in Figure~2 with the FastPM results
  \citet{Feng2016} (see their Figure~2).  For simulations with the
  same force resolution of $0.34\Mpch$, PPM-GLAM performs more
  accurately in spite of the fact that FastPM used Fourier-space to
  avoid the numerical differentiation of the gravitational
  potential.}. However, the computational cost of such a modification
is very substantial. It requires doubling the memory (additional
3D-mesh for accelerations) and also doubling the CPU time (3 inverse
FFTs instead of just one in our PPM-GLAM code).

\section{Effects of time-stepping and force-resolution}
\label{sec:AppendixB}
\subsection{Effects of time-stepping}
\makeatletter{}\begin{figure}
  \centering
\includegraphics[width=0.47\textwidth]
{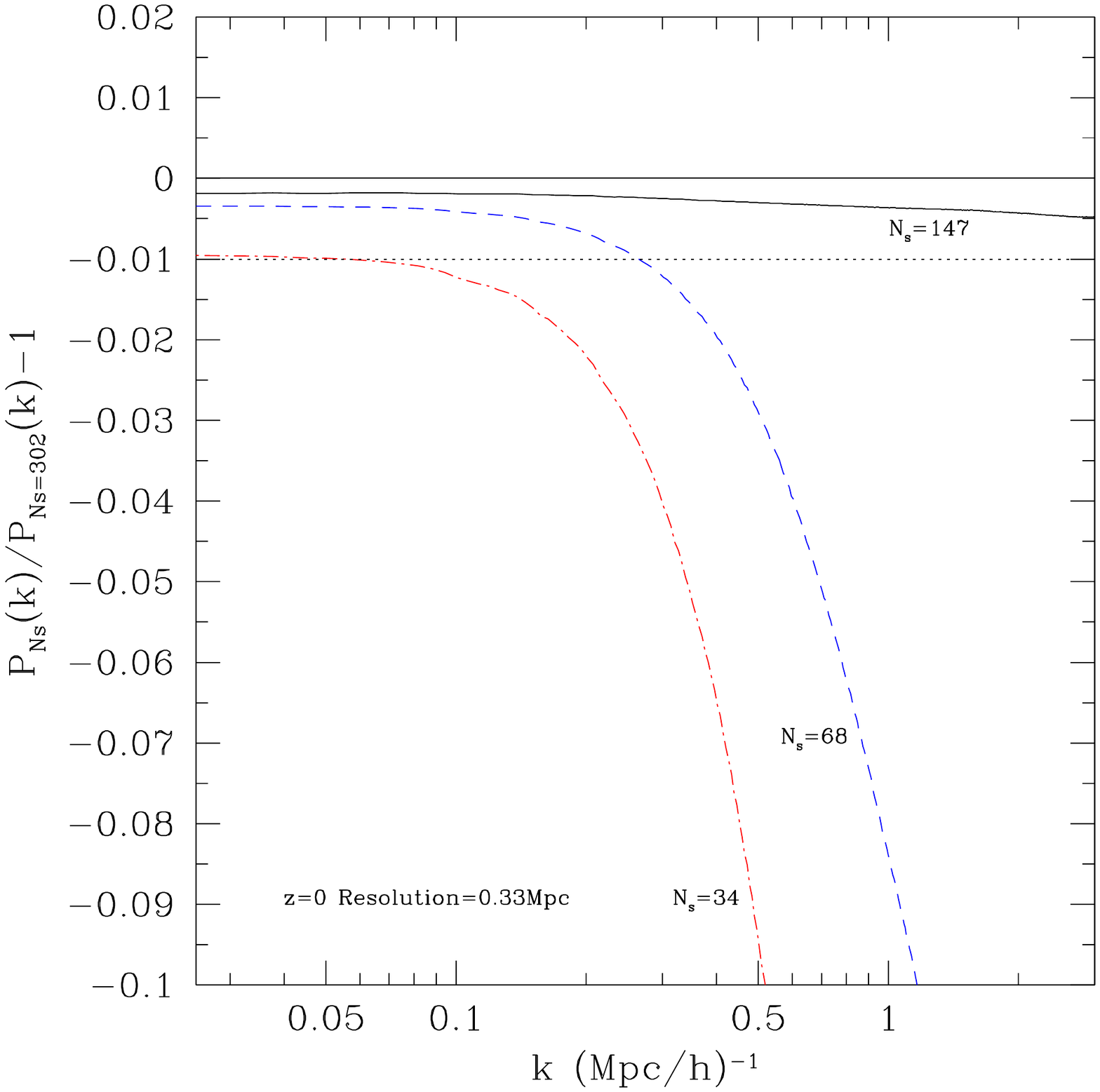}
\caption{Effects of the number of time-steps on the amplitude and convergence of the
  power spectrum. The simulations have the same initial conditions, number of particles
  $\Np=1000$, box size $1\Gpch$ and force resolution $\Delta x =0.33\Mpch$. The only difference 
  between $P(k)$ for various realisations, relative to the power spectrum of the simulation with the
largest number of time-steps $\Ns=302$, is the adopted number of time-steps, which is indicated in the plots.  
Results clearly converge when the number of steps increases and becomes
  $\gtrsim 100$ with very little difference between simulations with
  $\Ns=147$ and $\Ns=302$.}
\label{fig:ConvergeStepA}
\end{figure}

 \makeatletter{}\begin{figure}
  \centering
\includegraphics[width=0.47\textwidth]
{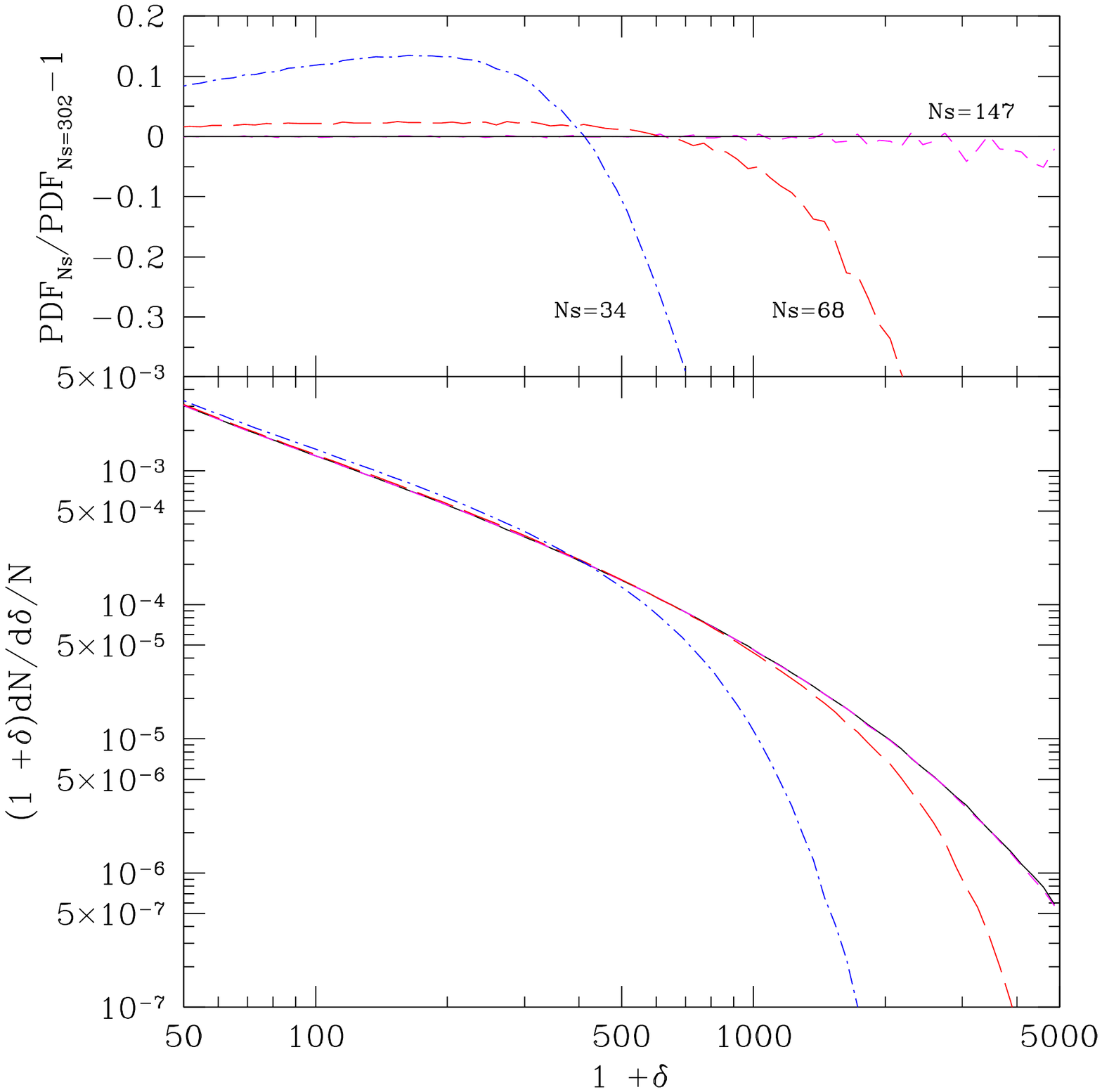}
\caption{The same as in Figure~\ref{fig:ConvergeStepA} but for the
 convergence of the density distribution function. Results clearly converge when the
  number of time-steps increases and becomes $\gtrsim 100$. However, a 
  smaller number of steps results in a dramatic suppression of the
  number of high-density regions where DM particles move very fast,
  which is observed as an artificial scale-dependent bias.}
\label{fig:ConvergeStepB}
\end{figure}

We have already presented in the main text various performance results
that indicate that the time-stepping adopted in our PPM-GLAM
simulations is adequate for accurate predictions of the dark matter
clustering statistics. Indeed, left panel in
Figure~\ref{fig:ConvergePk} compares the power spectrum in our
PPM-GLAM simulations with the MultiDark results. The latter
simulations have many more time-steps than the PPM-GLAM runs. There
are deviations but those are clearly related with the force resolution
and not the time-stepping. However, those comparisons are somewhat
convoluted because they test both the effects of force resolution and
time-stepping. Here we test the effects of time-stepping in a
different way.

We run a series of C1a simulations that start with the same initial
conditions at $z_{\rm init}=100$, have the same force resolution
$\Delta x=0.33\,\Mpch$, and differ only by the time-stepping
parameters.  Specifically, we run four realisations with box size $1\Gpch$,
$\Np=1000$ and $\Ng=3000$. The number of time-steps was changing
almost by a factor of two from one simulation to another with $\Ns = 34, 68, 147, 302$. 
The two runs with $\Ns = 34, 68$ have the
time-step $\Delta a/a \approx 0.15, 0.06$ all the time, while the
other two runs have $\Delta a/a$ at $z>3$ limited to
$\Delta a/a \approx 0.036, 0.015$ for $\Ns = 147, 302$ correspondingly,
and a constant $\Delta a$ at later moments. At $z=0$ they had
$\Delta a/a \approx 0.014, 0.006$ for $\Ns = 147, 302$
respectively. For comparison, a run with a constant $\Delta a$,
initial $z_{\rm init}=39$, and $\Ns=40$ has $\Delta a/a =0.024$ at
$z=0$ and $\Delta a/a \approx 0.1$ at $z=3$.

Figure~\ref{fig:ConvergeStepA} shows results for the power spectrum of
fluctuations relative to the power spectrum of the simulation with the
largest number of time-steps $\Ns=302$. There are clearly significant
errors in the simulations with the smaller number of steps. Note that
the errors are small at long waves which indicates that even a small
number of steps is sufficient for tracking the linear growth of
fluctuations. However, the errors dramatically increase at small scales because
the code cannot keep particles with large velocities inside dense
regions. When the number of steps increases the accuracy also improves
very substantially, we clearly see converge of the power spectrum when the
number of steps increases and becomes $\gtrsim 100$.

The power spectrum may give somewhat too optimistic impression. After
all, even 34 time-steps give an error in $P(k)$ of only 3\% at
$k\sim 0.3\,h{\rm Mpc}^{-1}$. The problem is that the error is much
larger if we look at dense regions. We study this effect by analyzing
the density distribution function of dark matter PDF, i.e. the fraction of volume
occupied by cells with a given overdensity $\delta$. In order to do that
we find the density in each cell of the $3000^3$ mesh, and count the
number of cells in a given density range $(\delta,\delta+\Delta\delta)$. 
Figure~\ref{fig:ConvergeStepB}, bottom panel, shows the PDF for those C1a simulations with 
different number of time-steps. At low densities, the PDF is
relatively insensitive to the number of steps, and this is why the errors in
$P(k)$ were also reasonable at long-waves. The situation is quite
different at large densities: relative errors are very large for densities
$\delta> 1000$, see top panel in Figure~\ref{fig:ConvergeStepB}. The plot also shows a
clear convergence for the simulations with the larger number of steps with very little difference
between $\Ns=147$ and $\Ns=302$.

\subsection{Effects of force resolution}
\makeatletter{}\begin{figure} \centering
\includegraphics[width=0.49\textwidth]
{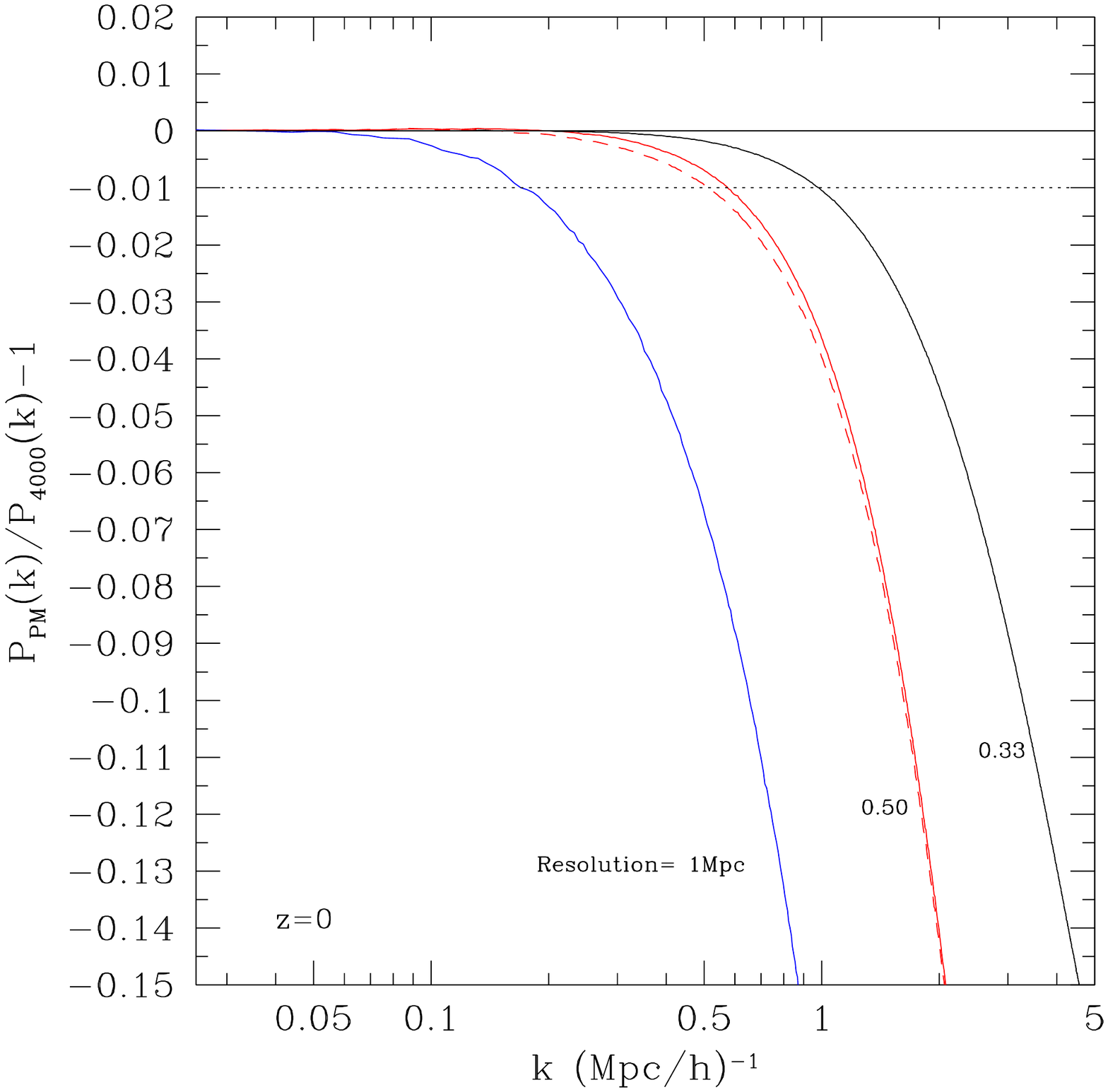}
\caption{Effects of force resolution on the power spectrum $P(k)$ at $z=0$ of a 
series of C1b simulations with the same number of
  particles $N_p=1000^3$ and computational box
  $L=1000\Mpch$. We run these simulations for grid sizes of $\Ng=$1000,
  2000, 3000, and 4000 with the force resolution ranging from
  $\Delta x= 0.25\Mpch$ to $1\Mpch$. The plot shows the ratio of the power spectrum
  $P(k)$ in each simulation to that set-up with the highest resolution run
  $\Ng=4000$. The dashed-curve is for a simulation with twice
  larger number of time-steps. With a resolution of $\Delta x= 0.5\Mpch$ the
  number of steps $\Ns\approx 100$ was sufficient.}
\label{fig:ConvergePMA}
\end{figure}

\makeatletter{}\begin{figure} \centering
\includegraphics[width=0.49\textwidth]
{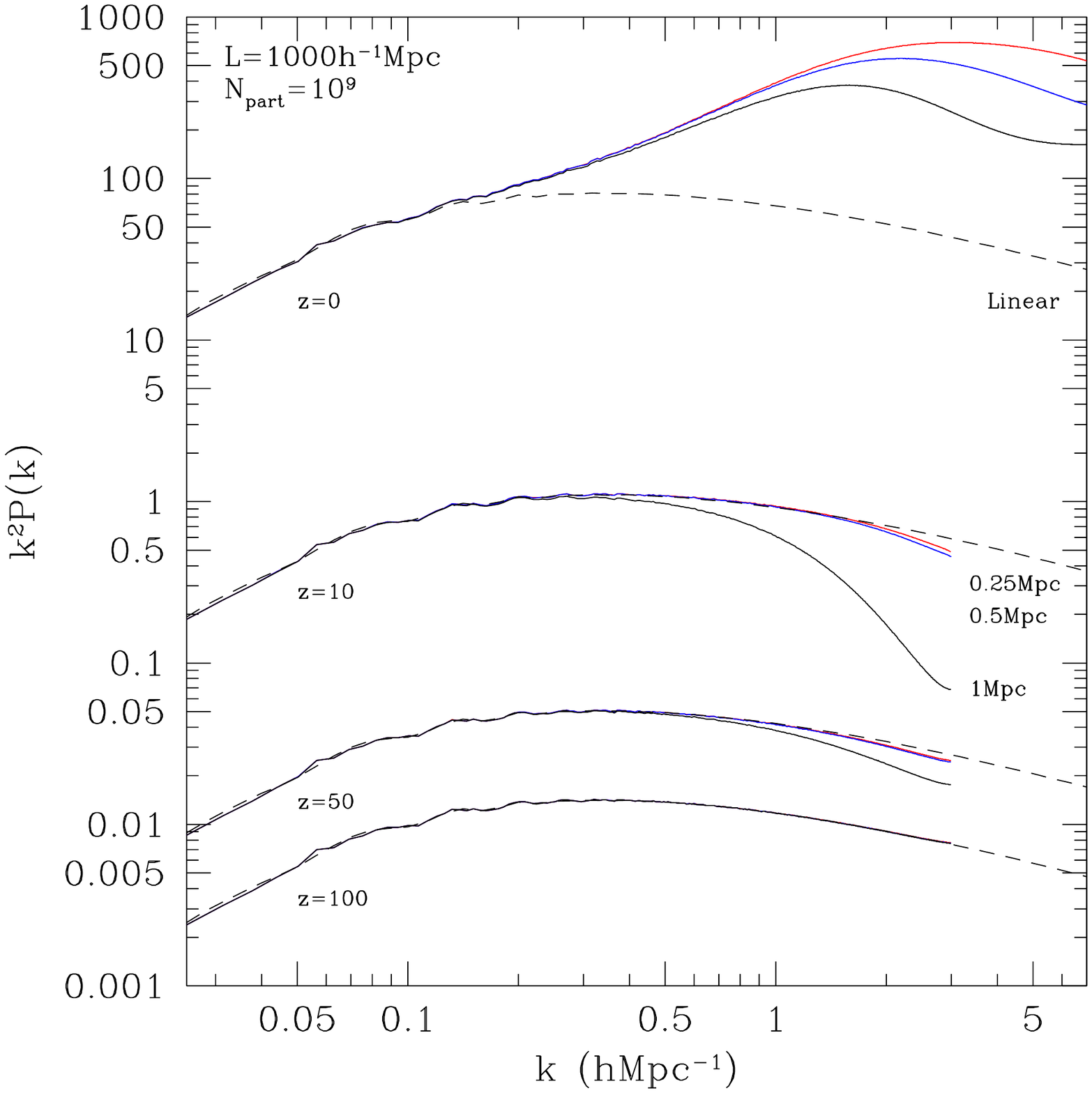}
\caption{Evolution of the
  power spectrum with redshift for the various C1b simulations. Better force resolution increases
  the power spectrum, but there are clear indications of convergence
  at a fixed wavenumber. The simulation with the number of particles
  equal to the mesh-size (labeled 1\,Mpc in the plot) shows
  disproportionally large suppression of fluctuations at initial
  stages of evolution.}
\label{fig:ConvergePMB}
\end{figure}

Figures~1 and 2 in the main text show how the power
spectrum converges as the force and mass resolution increase. Here we
present results of some additional tests. In order to study the effects of the
force resolution we run the C1b simulations with the same number of
time-steps $\Ns=136$ and number of particles $N_p=1000^3$, and change
the force resolution from $\Delta x= 0.25\,\Mpch$ to
$1\,\Mpch$ by running the same initial conditions using
different mesh-sizes. The smallest mesh has the same number of grid
points as the the number of particles $N_g=N_p=1000$. We then run other
simulations with $N_g=2000^3, 3000^3$, and $4000^3$ meshes. We also
run an additional simulation with $N_g=2000^3$ but with twice
larger time-steps ($\Ns=270$). Figure~\ref{fig:ConvergePMA} presents the ratio at $z=0$ of the
power spectrum $P(k)$ in each simulation to that with the highest
resolution $\Ng=4000$. 

Figure~\ref{fig:ConvergePMB} shows the evolution of the power spectra (scaled by $k^2$ to reduce the
dynamical range) in these C1b simulations. 
Results indicate $\sim 1\%$ convergence for
$k\lsim 1\,h{\rm Mpc}^{-1}$. At smaller scales the error increases,
but it is still $\sim 20-30\%$ even at
$k\approx (3-5)\,h{\rm Mpc}^{-1}$, which is also consistent with what
we find from the comparison with the MultiDark simulations in Figure~1
(left panel).

The evolution of the power spectra presented in 
Figure~\ref{fig:ConvergePMB} demonstrates significant suppression of
fluctuations for the simulation with the same number of particles and
mesh cells $N_g=N_p=1000$. 

%\makeatletter{}\begin{figure*}
%  \centering
%\includegraphics[width=0.59\textwidth]
%{powerTestZ100.pdf}
%\caption{Effects of force resolution on power spectrum $P(k)$. Plot
%  shows power spectra at different redshifts for simulations with the
%  same number of particles $N_p=1000^3$ and for computational box
%  $L=1000\Mpch$. Full curves are for grid sizes 1000, 2000, and 4000
%  (from bottom to top at each redshift). Dashed curves show
%  predictions of the linear theory. Initially (at $z=100$) all
%  simulations had the same random phases and power spectrum. Better
%  force resolution increases the power spectrum, but there are clear
%  indications of convergence at a fixed wavenumber. The simulation
%  with the number of particles equal to the mesh size (labeled $1Mpc$
%  in the plot) shows disproportionally large suppression of
%  fluctuations at initial stages of evolution. }
%\label{fig:ConvergeMesh}
%
%\end{figure*}

\end{document}